\documentclass[twocolumn]{aastex631}
\usepackage{graphicx}
\usepackage[T1]{fontenc}
\usepackage{amssymb}
\usepackage{amsmath}
\usepackage{CJK}

\newcommand{\revise}[1]{#1}

\begin{document}
\begin{CJK*}{UTF8}{gbsn}

\title{Global Simulations of Gravitational Instability in Protostellar Disks with Full Radiation Transport\\I. Stochastic Fragmentation with Optical-depth-dependent Rate and Universal Fragment Mass}

\author[0000-0002-9408-2857]{Wenrui Xu (许文睿)}
\affiliation{Center for Computational Astrophysics, Flatiron Institute, New York, NY 10010, USA}
\author[0000-0002-2624-3399]{Yan-Fei Jiang (姜燕飞)}
\affiliation{Center for Computational Astrophysics, Flatiron Institute, New York, NY 10010, USA}
\author[0000-0003-1676-6126]{Matthew W. Kunz}
\affiliation{Department of Astrophysical Sciences, Princeton University, Peyton Hall, Princeton, NJ 08544, USA}
\affiliation{Princeton Plasma Physics Laboratory, PO Box 451, Princeton, NJ 08543, USA}
\author[0000-0001-5603-1832]{James M. Stone}
\affiliation{Institute for Advanced Study, 1 Einstein Drive, Princeton, NJ, 08540, USA}

\begin{abstract}
Fragmentation in a gravitationally unstable accretion disk can be an important pathway for forming stellar/planetary companions. To  characterize quantitatively the condition for and outcome of fragmentation under realistic thermodynamics, we perform global 3D simulations of gravitationally unstable disks at various cooling rates and cooling types, including the first global simulations of gravitational instability that employ full radiation transport.
We find that fragmentation is a stochastic process, with the fragment generation rate per disk area $p_{\rm frag}$ showing an exponential dependence on the parameter  $\beta\equiv\Omega_{\rm K} t_{\rm cool}$, where $\Omega_{\rm K}$ is the Keplerian rotation frequency and $t_{\rm cool}$ is the average cooling timescale. Compared to a prescribed constant $\beta$, radiative cooling in the optically thin/thick regime makes $p_{\rm frag}$ decrease slower/faster in $\beta$; the critical $\beta$ corresponding to $\sim 1$ fragment per orbit is $\approx$3, 5, 2 for constant $\beta$, optically thin, and optically thick cooling, respectively.
The distribution function of the initial fragment mass is remarkably insensitive to disk thermodynamics. Regardless of cooling rate and optical depth, the typical initial fragment mass is $m_{\rm frag} \approx 40 M_{\rm tot}h^3$, with $M_{\rm tot}$ being the total (star+disk) mass and $h=H/R$ being the disk aspect ratio.
Applying this result to typical Class 0/I protostellar disks, we find $m_{\rm frag}\sim 20 M_{\rm J}$, suggesting that fragmentation more likely forms brown dwarfs. Given the finite width of the $m_{\rm frag}$ distribution, forming massive planets is also possible.
\end{abstract}

\section{Introduction}
\subsection{Gravitational instability in protostellar disks}

Gravitational instability (GI) occurs in accretion disks when the disk has sufficiently high surface density and low temperature so that self-gravity becomes dynamically important.
GI may occur in astrophysical systems such as galactic disks, AGN disks, and protostellar/protoplanetary disks \citep[see reviews by][]{Lodato07,Shu16,KratterLodato16}. The basic theory of GI is universally applicable to all these contexts, but the detailed physics varies;
here we focus our attention on protostellar disks.

Theoretically, GI may be common among young protostellar disks. In Class 0/I (and possibly young Class II) protostellar systems, GI may be triggered by the high influx of mass provided by the infalling envelope or (in the inner disk) by the accretion mass flux from the outer disk \citep{LinPringle87,AdamsLin93}.
However, when the magnetic field and non-ideal MHD effects are taken into consideration, simulations of protostellar collapse and disk formation do not yet agree on whether GI should be prevented by magnetic angular-momentum transport and removal (see the recent review by \citealt{Tsukamoto+23}; also see discussions in \citealt{XK21a,XK21b}).


Observationally, there is a growing sample of individual disks that show some evidence of GI \citep[e.g.,][]{Huang+18,Huang+20,Lee+20,Paneque-Carreno+21,Veronesi+21,Xu+23,Lodato+23,Burns+23}. Meanwhile, surveys of Class 0/I systems estimate a typical disk mass of $0.01$--$0.1~{\rm M}_\odot$ \citep{Tychoniec+18,Andersen+19,Tobin+20}, which is comparable to or lower than what is typically required for GI. However, these estimates assume optically thin emission, which can significantly underestimate the disk mass \citep{GalvanMadrid+18,Liu20}. Indeed, studies using more optically thin wavelengths \citep[e.g.,][]{Tychoniec+18,Sharma+20} or tracers \citep[e.g.,][]{Booth+19,BoothIlee20} find higher disk masses that are susceptible to GI. Additionally, semi-analytic models of Class 0/I disks self-regulated by GI  reproduce both the observed fluxes and the radial intensity profiles and predict other observables such as the typical dust-continuum spectral index \citep{Xu22,Xu+23}.

The effect of GI on disk evolution and planet formation can be divided into two categories: fragmentation and gravitoturbulence.
With sufficiently strong cooling, GI results in fragmentation, which is a promising (though not definitive) explanation for the formation of stellar/planetary companions observed at ${\sim}100$~au separations \citep{NeroBjorkman09,Kratter+10,Nielsen+19,Vigan+21,Offner+23}. More generally, GI regulates disk evolution via angular-momentum transport and heating produced by spirals and clumps (``gravitoturbulence'').
Gravitoturbulence, when cascaded down to small scales, can regulate the collisional coagulation of dust grains \citep{XA23}. The pressure maxima associated with spirals and clumps may also concentrate larger dust grains and promote planetesimal/embryo formation \citep{Rice+04,Rice+06,Gibbons+12,Gibbons+14,Baehr+22,Rowther+24}.

\subsection{Existing studies and key uncertainties}

For both fragmentation and gravitoturbulence, existing studies have provided a basic picture, but there remain a few important uncertainties that limit our ability to model GI accurately in theoretical and observational contexts.

In terms of predicting fragmentation, some simulations find that fragmentation occurs when the dimensionless parameter  $\beta\equiv\Omega_{\rm K} t_{\rm cool}\lesssim 3$, where $\Omega_{\rm K}$ is the Keplerian frequency and $t_{\rm cool}$ is the average cooling timescale \citep{Gammie01,Deng+17,Baehr+17}.
However, this threshold does not capture the whole picture, because fragmentation is likely a stochastic process: disks with larger $\beta$ may still fragment, though at a much reduced rate (\citealt{Paardekooper+11}; also see \citealt{HopkinsChristiansen13,Brucy+21}).
Predicting fragmentation thus requires a quantitative prescription for the fragment generation rate per disk area as a function of $\beta$ (and possibly other disk parameters).

The distribution of initial fragment mass also remains somewhat uncertain. Dimensional analysis suggests that the initial fragment mass should scale with $\Sigma H^2$ (with $\Sigma$ being the surface density and $H$ the disk scale height). However, what is the prefactor for this scaling, and does it depend upon dimensionless parameters such as $\beta$? Analytic estimates of this prefactor differ by an order of magnitude \citep[see Section~5.1 of][]{KratterLodato16}. Meanwhile, results from 2D (thin-disk) simulations \citep[e.g.,][]{Zhu+12,Vorobyov13} are inevitably affected by the 2D setup, because realistic fragments form at scales ${\lesssim}H$. Finally, some 3D simulations include fragment mass measurements \citep[e.g.,][]{Hall+17,MercerStamatellos17,Deng+21,Boss21,BossKanodia23,Kubli+23}, but they do not  calibrate quantitatively how the characteristic fragment mass scales with disk parameters such as $\Sigma, H$, and $\beta$. As a result, it is difficult to generalize these results beyond their particular physical setups.

In terms of predicting the transport produced by the spirals and clumps in gravitoturbulence and the observational signatures of these substructures, there is a tension between two different and seemingly incompatible paradigms.
\revise{In the context of modeling spiral substructures, especially for observational purposes, a common assumption is that these substructures are dominated by a single global linear eigenmode of GI \citep{LinShu66}, and this assumption seems consistent with some observations \citep[e.g.,][]{Paneque-Carreno+21,Xu+23}. In the context of modeling transport, however, GI is often modeled using a local  viscosity \citep{LinPringle87,Gammie01}, and it can be argued analytically that this approximation is reasonable only when perturbations are localized around corotation \citep{BalbusPapaloizou99}. One possible reconciliation is that these two paradigms may apply to different disk masses. A general trend among simulations is that lower disk-to-mass ratio leads to more localized perturbations with time-averaged transport broadly consistent with a local viscosity \citep[e.g.,][]{LodatoRice04,Cossins+09,XK21b,Bethune+21,Steiman-Cameron+23}, and higher disk-to-mass ratio leads to more global spirals and the transport no longer represents a local viscosity \citep[e.g.,][]{LaughlinRozyczka96,LodatoRice05}. But this is not a satisfactory answer; even for the simulations with more massive disks showing more global spirals, these spirals lack the $m$-fold symmetry characteristic of linear eigenmodes. It is also unclear whether the difference between low-mass and high-mass disks are due to fundamental difference in the dynamics, or they share the same dynamics but with the typical scale of perturbation and the deviation from local viscosity proportional to the scale height $H$. (For a marginally unstable disk, the aspect ratio $H/R$ scales with the disk-to-star mass ratio.)}


Finally, the role of realistic thermodynamics in determining the fundamental properties of GI (and how they should be modeled) remains ambiguous. A large fraction of existing studies prescribe a constant cooling time $\beta$.
Radiative cooling in real disks, however, produces a temperature-dependent $\beta$ and is qualitatively different in optically thin (streaming) and optically thick (diffusing) regimes.\footnote{The optically thick regime can be further divided according to whether the light diffusion time is longer or shorter than the dynamical timescale. For protostellar disks, the dynamics is non-relativistic and the optical depth is not too high ($\tau\lesssim 10^3$; see \citealt{Xu22}), so we consider only the case of short light diffusion time.}
Radiative cooling has been modeled, under various degrees of approximation, in a number of studies \citep[e.g.,][]{Boley+06,HiroseShi19,Boss21,XK21b,Steiman-Cameron+23}. The existing works, however, do not include a systematic comparison between constant-$\beta$ simulations and radiative simulations, so it is often unclear which results  depend only on the overall cooling rate (effective $\beta$) and which results depend on the cooling type.
Moreover, most existing studies do not  solve directly the full radiation transport equations for the direction-dependent intensity field, but rather employ approximations such as optical-depth-based parametrization and/or flux-limited diffusion. These approximations introduce order-unity errors in marginally optically thick ($\tau\sim 1$) regions, which are often where most of the cooling takes place.

\subsection{Our work}

We address the key uncertainties summarized in the previous subsection through a set of global 3D simulations of GI. We systematically survey different cooling rates and cooling types, covering constant-$\beta$ cooling and radiative cooling in both optically thin and optically thick regimes. We use our simulation results to clarify the physical picture of GI and provide quantitative prescriptions for modeling GI in semi-analytic models. Given the wide scope of this study, we divide our results into two papers. Paper I (this paper) introduces the simulation setup and presents results on fragmentation. Paper II \citep{paper2} focuses on spirals/clumps in gravitoturbulence and the transport and observational signatures they produce.

The rest of this paper is organized as follows.
In Section~\ref{sec:setup}, we introduce the simulation setup, including how we define fragmentation. In Section~\ref{sec:results}, we summarize the simulation results and measure the rate of fragmentation and the fragment masses. We discuss the physical picture of fragmentation in Section~\ref{sec:discussion}, and the implications of our results for planet/binary formation via fragmentation in protostellar disks in Section~\ref{sec:application}. We summarize our findings in Section~\ref{sec:conclusion}.

\section{Simulation setup}\label{sec:setup}
\subsection{Code and units}
We perform global hydrodynamic and radiation hydrodynamic simulations using \texttt{Athena++} \citep{Stone+20}. Self-gravity of the disk is included by solving the discrete Poisson equation using the algorithm described in \citet{XK21a}. In simulations with radiation transport, we solve the full radiation transport equations using the algorithm described in \citet{Jiang21,Jiang22}. This algorithm is implicit, so the fast speed of light does not limit the timestep of the simulation.

We choose the unit mass $m_0$ to be the total (combined) mass of the star and the disk $M_{\rm tot}$, the unit length $r_0$ to be the initial disk size, and the unit time to be $t_0 \equiv \Omega_0^{-1} \equiv (GM_{\rm tot}/r_0^3)^{1/2}$. Under this normalization, the gravitational constant $G$ is unity.

\subsection{Domain and resolution}\label{sec:setup:res}
All simulations adopt a spherical-polar domain with $r\in[0.25r_0,2r_0]$, $\theta\in[0,\pi/2]$, and $\phi\in[0,2\pi]$. A reflecting boundary condition is used at the midplane ($\theta=\pi/2$). Some of our definitions and diagnostics require cylindrical coordinates, and we use $(r,\theta,\phi)$ and $(R,\phi,z)$ to denote spherical and cylindrical coordinates, respectively.

We use a log-uniform $r$ grid, a uniform $\phi$ grid, and a non-uniform $\theta$ grid that has its highest resolution near the midplane and much lower resolution around the pole. The $\theta$ grid consists of two regions: for $\theta\in[\pi/2-h_0, \pi/2]$ with $h_0=0.05$ ($\approx 1$ scale height for our disk), the resolution is uniform. For $\theta \in [0,\pi/2-h_0]$, the resolution decreases towards the pole, with a fixed ratio between the sizes of neighboring cells. The size ratio is chosen such that the cell size is continuous across $\pi/2-h_0$.

Our default resolution is $(N_r, N_\theta, N_\phi) = (480,32,480)$ with the high-resolution $\theta$ grid (0.05 rad) containing 16 cells. This gives a resolution of $\approx (1/12, 1/16, 1/4)$ scale height near the midplane. We also perform low-resolution runs with $(N_r, N_\theta, N_\phi) = (240,24,240)$, with the high-resolution $\theta$ grid contains 8 cells. Near the midplane, this gives half of our fiducial resolution in all directions.

In runs with radiation transport, the angular grid for specific intensity $I$ consists of 32 directions. This angular grid is defined in a polar coordinate system with the pole aligned with the radial direction of our spatial grid; it is spanned by 4 angles between $[0,\pi]$ along the polar direction and 8 angles between $[0,2\pi]$ along the azimuthal direction.

\subsection{Initial condition}

We use a gravitoturbulent disk as the initial condition of all our simulations.
Compared to starting from a non-turbulent disk, this reduces spurious fragmentation \citep{Paardekooper+11}. 
\revise{This gravitoturbulent state is produced by evolving a marginally unstable disk with low-amplitude seed perturbation under a modest linear cooling rate $\beta\equiv\Omega_{\rm K}t_{\rm cool}=10$ for $30 t_0$. Below we detail how we prepare the marginally unstable disk and the seed perturbation.}

We start from a Keplerian torus with rotation rate
\begin{equation}
    \Omega_{\rm K} \equiv \sqrt{\frac{GM_{\rm tot}}{R^3}}.
\end{equation}
The initial surface density of the torus is
\begin{equation}
    \Sigma(R) = \begin{cases}
\Sigma_{\rm init}(R) & {\rm for}~R_{\rm d,in}\leq R\leq R_{\rm d,out},\\
0 &\text{otherwise,}
\end{cases}
\end{equation}
with
\begin{equation}
    \Sigma_{\rm init}(R) \equiv \frac{M_{\rm d}}{2\pi R^2 \ln(R_{\rm d,out}/R_{\rm d,in})} \propto R^{-2}.
\end{equation}
Here the disk mass $M_{\rm d}$ is chosen to be $0.1 M_{\rm tot}$ (similar to previous studies, e.g., \citealt{Cossins+09,Deng+17,Steiman-Cameron+23}), and the inner and outer boundaries of the initial torus are $R_{\rm d,in}=r_0/3$ and $R_{\rm d,out}=r_0$. We leave some space between the initial torus and the radial boundaries to reduce contamination due to wave reflections at the boundaries.
The $R^{-2}$ surface density profile we choose is motivated by the slope of gravitationally self-regulated disks seen in radiative simulations, semi-analytic modeling, and observation \citep[e.g.,][]{XK21b,Xu22,Xu+23}.\footnote{Due to this steep $\Sigma$ profile, the total mass of gravitationally self-regulated disks can be much larger than the local disk mass $\pi R^2\Sigma \sim 0.1 M_{\rm tot}$ \citep{Xu+23}. Since we only simulate a small radial range, our $M_{\rm d}$ remains similar the local disk mass $\pi R^2\Sigma$. Meanwhile, with large $R_{\rm d,out}/R_{\rm d,in}$, choosing a higher disk mass would be more appropriate \citep[see an example in the setup of][]{Bethune+21}.} We choose an initial temperature profile of $T\propto R^{-1}$, giving nearly uniform aspect ratio and uniform ``Keplerian Toomre $Q$'',
\begin{equation}
    Q_{\rm K} \equiv \frac{c_{\rm s}\Omega_K}{\pi G \Sigma}.
\end{equation}
Here $c_{\rm s}$ is the adiabatic sound speed. 
$Q_{\rm K}$ differs from the actual Toomre $Q$ in that the epicyclic frequency $\kappa$ is replaced with the Keplerian frequency $\Omega_{\rm K}$. In practice, it is hard to obtain reliable measurements of $\kappa$ in a highly turbulent, self-gravitating disk, so we use $Q_{\rm K}$ to describe the level of gravitational instability instead. Given our relatively low disk mass, this is usually a good approximation. For our initial condition, the temperature is chosen to give $Q_{\rm K} = 1.5$. We set the initial vertical density profile of the torus to be
\begin{equation}
    \rho(R,z) \propto {\rm e}^{-z^2/2H^2}
\end{equation}
with $H$ being the scale height defined by
\begin{equation}
    H\equiv \frac{c_{\rm s,iso}}{\Omega_{\rm K}}.
\end{equation}
Here $c_{\rm s,iso} = \gamma^{-1/2}c_{\rm s}$ is the isothermal sound speed. For all our simulations we adopt an adiabatic index of $\gamma=5/3$. \revise{(The choice of $\gamma$ also affects the fragmentation threshold; see \citealt{Rice+05}.)} This profile correspond to a vertical hydrostatic equilibrium when disk self-gravity is negligible. For our initial temperature profile, the initial aspect ratio is $H/R=0.053$.

\revise{We then introduce a small-amplitude azimuthal density perturbation to break the azimuthal symmetry. The perturbation consists of $m=1$--6 modes;}
for each mode, the phase and relative amplitude are constant across the disk, and they are chosen randomly from $[0,2\pi]$ and $[0,0.01]$.

\subsection{Cooling and radiation}\label{sec:setup:cooling}
We perform simulations with three different types of cooling/radiation prescriptions. Below we describe each prescription and the associated free parameters.

\textbf{Constant $\beta$ cooling:} We use a fixed cooling time at each radius, $t_{\rm cool}$, specified by a constant normalized cooling time
\begin{equation}
    \beta\equiv \Omega_{\rm K} t_{\rm cool}.
\end{equation}
This prescription is not realistic, but it is commonly adopted \cite[e.g.,][]{Gammie01} due to its simplicity. Our constant $\beta$ runs serve as a baseline for comparison with previous studies and with our other cooling prescriptions.

\textbf{Optically thin cooling:} In the limit where the disk is optically thin and the background radiation is negligible, the cooling rate per volume is
\begin{equation}
    q_{\tau\to 0} = -4\rho\kappa\sigma_{\rm SB}T^4,\label{eq:cooling_thin}
\end{equation}
with $\rho$, $\kappa$, and  $\sigma_{\rm SB}$ being the mass density, opacity, and Stefan-Boltzmann constant, respectively.
For simplicity, we assume the opacity $\kappa$ to be constant for each simulation. The cooling rate thus depends only on the value of $\kappa\sigma_{\rm SB}$ in code unit, which we set with the dimensionless parameter
\begin{equation}
    \beta_0 \equiv \frac{\Omega_0u_{\rm d}}{4\kappa\sigma_{\rm SB}T_{\rm d}^4}.
\end{equation}
Here $u_{\rm d}$ and $T_{\rm d}$ are the specific internal energy and temperature of our $Q_{\rm K}=1.5$ torus initial condition evaluated at $r_0$. $\beta_0$ thus corresponds to a rough estimate of the normalized cooling timescale $\beta$ of the disk.

\textbf{Optically thick cooling with full radiation transport:} More generally, the disk can have finite optical depth. In this case, we solve self-consistently for the radiation field and apply the corresponding energy and momentum source terms following \citet{Jiang21}. (For the regime considered in this paper, the momentum source terms due to radiation pressure remain small.) The evolution depends on $\kappa$, $\sigma_{\rm SB}$, and the speed of light $c$. For simplicity, we consider an opacity $\kappa$ that is wavelength independent and constant for each simulation, and we do not include scattering. We set (the code-unit value of) $\kappa$, $\sigma_{\rm SB}$, and $c$ by specifying the following three dimensionless parameters,
\begin{align}
    &\tau_0 \equiv \Sigma_{\rm d}\kappa,\\
    &\beta_0 \equiv \frac{\Omega_0 U_{\rm d}}{\Lambda_{\rm a}(\tau_0/2, T_{\rm d})},\\
    &c_0 \equiv \frac{c}{\Omega_0r_0}.
\end{align}
Here $\Sigma_{\rm d}, U_{\rm d}$, and $T_{\rm d}$ are the surface density, vertically integrated thermal energy, and temperature of our $Q_{\rm K}=1.5$ torus initial condition evaluated at $r_0$. $\Lambda_{\rm a}$ is an analytic estimate of the cooling rate per disk area as a function of the midplane optical depth $\tau_{\rm mid}$ and vertically averaged temperature $T_{\rm mean}$, given by
\begin{equation}
    \Lambda_{\rm a}(\tau_{\rm mid}, T_{\rm mean}) \equiv \frac{8\sigma_{\rm SB}\tau_{\rm mid}T_{\rm mean}^4}{[1+(0.875\tau_{\rm mid}^2)^{0.45}]^{1/0.45}}.\label{eq:cooling_analytic}
\end{equation}
The origin of this analytic estimate is discussed in Appendix \ref{a:cooling_analytic}.
Physically, $\tau_0$ and $\beta_0$ estimate the optical depth and normalized cooling time of the disk.
In the limit of $\tau_0\to 0$ and $c_0\to\infty$, full radiation transport is equivalent to optically thin cooling with the same $\beta_0$.

Because full radiation transport is expensive, we cannot survey a large parameter space. We fix $\tau_0=10$ and $c_0=10^5$ for all our radiation runs. The choice of $\tau_0=10$ covers the regime of optically thick disks; together with the optically thin cooling runs which represent $\tau_0\to 0$ disks, these two sets of runs cover the limiting cases of disk thermodynamics. With the light crossing and diffusion times being much faster than the dynamical timescale, radiation transport is effectively time-independent and the exact choice of $c_0$ should not significantly affect the evolution. Here we choose $c_0=10^5$ because $m_0=1~{\rm M}_\odot$ and $r_0=100$~au correspond to $c_0\approx 10^5$.

\revise{In all three cases, our setup is essentially scale free as the dynamics is controlled by dimensionless parameters (e.g., $\tau_0,\beta_0$). This allows our results to be easily rescaled to parameters of specific protostellar disks (or disk populations) and provide more accurate estimates (Section \ref{sec:application:frag_mass}). Our results may also be generalized to other gas-dominated gravitationally unstable disks, such as gas-dominated regions in AGN disks. Meanwhile, a limitation of our choice is that we only consider a constant opacity. Some previous studies have used more realistic opacity tables that include the temperature-dependent dust opacity of protostellar disks \citep[e.g.,][]{MercerStamatellos17,Meru+17}, and we discuss the possible effect of a temperature-dependent opacity in Section \ref{sec:discussion:opacity}.

For our radiative simulations, we do not include the irradiation from the central protostar. This is mainly because protostellar disks self-regulated by GI are usually self-shadowing, with the innermost part of the disk having a larger geometric thickness and thus shielding the rest of the disk from direct stellar irradiation \citep[e.g.,][]{XK21b,Xu+23}. This choice also facilitates a more direct comparison with constant $\beta$ simulations. Meanwhile, shearing-box simulations that include stellar irradiation by \citet{HiroseShi19} find that the GI dynamics depend on the average $\beta$ in a way similar to non-irradited simulations.}

\subsection{Boundary conditions and accretion}
For the $\theta$ and $\phi$ directions, we adopt the appropriate geometric boundary conditions, which are polar at $\theta=0$, reflecting at $\theta=\pi/2$, and periodic at $\phi=0$ and $2\pi$.

For the inner radial boundary, we use a modified outflow boundary where $\rho, P, r^2v_r, v_\theta, r^{-1}v_\phi$ and the specific intensity $I$ have zero gradient across the boundary. We also cap the radial velocity in ghost cells to $v_r\leq 0$, so that there is no inflow into the domain. Mass that flows across the inner boundary is added to the central point mass gravitational potential representing the star. For our simulations, the stellar mass changes by less than $1\%$. One caveat is that we always assume the star to be fixed at the origin when computing its potential, and this could produce some small error when the disk contains asymmetric ($m=1$) density perturbations.

For the outer radial boundary, we let $\rho, P, v_\theta, r^{-1}v_\phi$ have zero gradient across the boundary. We impose a reflecting boundary condition on $r^2v_r$ to avoid mass flow through the outer boundary. A vacuum boundary condition is used for radiation, where the outgoing components of $I$ have zero gradient across the boundary and the incoming components of $I$ are fixed at zero in the ghost cells.

\subsection{Fragments and fragmentation}

\subsubsection{Defining and identifying fragments}
For this study, we use the term ``clump'' to refer to general overdensities that may or may not be gravitationally bound, and use ``fragment'' to  refer exclusively to gravitationally bound clumps, defined as follows.


First, in each simulation snapshot we search for gravitational potential wells formed by dense clumps. With respect to a given radius $R_1$, the effective gravitational potential in a frame corotating at a velocity that balances the gravity at $R_1$ is defined by
\begin{equation}
    \Phi_{\rm eff}(R,\phi,z,R_1) = \Phi(R,\phi,z) - \frac 12\Omega_g^2(R_1)R^2,
\end{equation}
where $\Phi$ is the gravitational potential (including both the star and the disk), and $\Omega_g^2(R_1)\equiv -g_r(R_1)R_1$ with $g_r$ being the gravitational acceleration corresponding to the azimuthally averaged midplane gravitational potential. Using this effective potential accounts for the tidal force in the rotating frame. If a point $(R,\phi,z)$ is the local minima of $\Phi_{\rm eff}(R,\phi,z,R_1)$ when $R_1=R$, we mark this point as the center of a potential well.

For each potential well center, we then consider all closed isosurfaces of $\Phi_{\rm eff}$ that contains it. Here ``closed'' means that the isosurface does not contain the inner or outer boundary of the domain. For each closed isosurface, consider the total energy within it,
\begin{equation}
    E_{\rm tot} = \int_C \left[\frac{P}{\gamma-1} + \rho|\boldsymbol{v}-\boldsymbol{v}_{\rm mean}|^2 + \rho(\Phi_{\rm eff}-\Phi_{\rm eff, C}) \right]{\rm d}V.
\end{equation}
Here $C$ marks the region bound by the isosurface, $\boldsymbol{v}_{\rm mean}$ is the density-weighted average velocity in $C$, and $\Phi_{{\rm eff}, C}$ is $\Phi_{\rm eff}$ on the isosurface. If there exists any closed isosurface that is gravitationally bound ($E_{\rm tot}<0$), we use the largest closed isosurface that is gravitationally bound to define a fragment. Material within this largest closed and bound isosurface is counted as the fragment mass.

\subsubsection{Tracing fragments}
In general, a fragment can live across multiple snapshots; in this case we should count them as the same fragment to avoid overestimating the fragment generation rate. To match the same fragment between two adjacent snapshots, we azimuthally advect each fragment in the first snapshot by $\Delta\phi = \Omega_{\rm K}\Delta t$ to predict its location in the second snapshot. Here $\Delta t$ is the interval between adjacent snapshots; for our simulations, $\Delta t = 0.05 t_0$.
If the predicted location is close to a fragment in the second snapshot (separation $<20\%$ of advected distance; the result is not sensitive to the exact choice of this threshold), we label these two fragments as the same; otherwise, we record that the fragment has been disrupted before the second snapshot. One caveat is that, if a clump repeatedly oscillates between being bound and unbound, it can be incorrectly counted as multiple fragments; tracking the time evolution of the ratio between the total and gravitational energy of individual clumps shows that such behavior is uncommon in our simulations. 

\subsubsection{Numerical fragment disruption}\label{sec:fragment_disruption}
One key limitation of our numerical setup is that we cannot reliably trace the gravitational collapse of fragments beyond the first few $\Omega_{\rm K}^{-1}$ due to the fixed spatial resolution. Once a fragment collapses down to the grid scale (i.e., once most of the fragment mass is contained within a few cells), it tends to produce some unphysical energy injection that unbounds and dissolves the fragment (see Appendix~\ref{a:disruption}). As a result, a fragment that would continue to collapse and survive for an extended period of time in the limit of infinite resolution is disrupted over several $\Omega_{\rm K}^{-1}$ in our simulation. This makes our setup not suitable for studying the long-term evolution of fragments. 

On the other hand, the numerical disruption of fragments allows each simulation to cover the generation of many fragments. 
A collapsing fragment can slow down (or even break) the simulation by introducing extreme velocities and temperatures, in which case the simulation would barely proceed after the first collapsing fragment forms. But since bound fragments always dissolve before reaching extreme densities in our simulations, we are able to simulate a fragmenting disk for an extended period of time and cover the generation of a large number of fragments. With this, we can obtain more reliable statistics on the fragment generation rate and the fragment initial mass function.
We also comment that in our simulations, the numerical heating associated with a single fragment never injects so much energy that it affects the global properties of the disk or artificially stops it from further fragmentation.

\begin{deluxetable*}{lccccccccc}
\tablecaption{Summary of simulations.\label{tab:runs}}
\tablenum{1}
\tablehead{
\colhead{Name} &
\colhead{$T/t_0$} &
\colhead{cooling type} &
\colhead{$\beta_0$} &
\colhead{$\tau_0$} &
\colhead{$c_0$} &
\colhead{$\beta$} &
\colhead{$\beta_1$} &
\colhead{$n_{\rm frag}$} &
\colhead{$f_{\rm frag}/t_0^{-1}$}
}
\startdata
\multicolumn{10}{c}{fiducial-resolution runs}\\
\hline
B2 & 30.0 & constant $\beta$ & - & - & - & 2 & - & 1.36 & 6.13 \\
B3 & 30.0 & constant $\beta$ & - & - & - & 3 & - & 0.24 & 1.43 \\
B3.5 & 30.0 & constant $\beta$ & - & - & - & 3.5 & - & 0.18 & 0.97 \\
B5 & 30.0 & constant $\beta$ & - & - & - & 5 & - & 0.03 & 0.20 \\
B10 & 30.0 & constant $\beta$ & - & - & - & 10 & - & 0 & 0 \\
T0\_B2 & 30.0 & optically thin & 2 & - & - & 2.48 & 2.28 & 0.83 & 3.57 \\
T0\_B3 & 30.0 & optically thin & 3 & - & - & 3.35 & 3.13 & 0.52 & 2.03 \\
T0\_B5 & 30.0 & optically thin & 5 & - & - & 5.49 & 4.44 & 0.13 & 0.77 \\
T0\_B10 & 30.0 & optically thin & 10 & - & - & 7.10 & 5.77 & 0.06 & 0.30 \\
T10\_B0.03 & 16.6\tablenotemark{a} & optically thick & 0.03 & 10 & $10^5$ & 2.21 & 1.00 & 0.44 & 1.26 \\
T10\_B0.1 & 30.0 & optically thick & 0.1 & 10 & $10^5$ & 4.72 & 2.04 & 0 & 0 \\
T10\_B0.3 & 30.0 & optically thick & 0.3 & 10 & $10^5$ & 7.57 & 2.90 & 0 & 0 \\
\hline
\multicolumn{10}{c}{low-resolution runs}\\
\hline
B2\_L & 30.0 & constant $\beta$ & - & - & - & 2 & - & 1.00 & 6.00 \\
B2.5\_L & 30.0 & constant $\beta$ & - & - & - & 2.5 & - & 0.49 & 3.17 \\
B3\_L & 30.0 & constant $\beta$ & - & - & - & 3 & - & 0.20 & 1.27 \\
B3.5\_L & 30.0 & constant $\beta$ & - & - & - & 3.5 & - & 0.09 & 0.60 \\
B5\_L & 30.0 & constant $\beta$ & - & - & - & 5 & - & 0.02 & 0.20 \\
B10\_L & 30.0 & constant $\beta$ & - & - & - & 10 & - & 0.01 & 0.03 \\
T0\_B2\_L & 25.0\tablenotemark{b} & optically thin & 2 & - & - & 1.20 & 1.29 & 1.10 & 6.64 \\
T0\_B3\_L & 30.0 & optically thin & 3 & - & - & 3.82 & 3.27 & 0.14 & 1.03 \\
T0\_B4\_L & 30.0 & optically thin & 4 & - & - & 6.55 & 5.59 & 0.02 & 0.10 \\
T0\_B5\_L & 30.0 & optically thin & 5 & - & - & 7.64 & 6.16 & 0 & 0 \\
T0\_B10\_L & 30.0 & optically thin & 10 & - & - & 11.41 & 7.70 & 0 & 0 \\
T10\_B0.01\_L & 8.0\tablenotemark{b} & optically thick & 0.01 & 10 & $10^5$ & 1.73 & 0.70 & 1.68 & 14.99 \\
T10\_B0.03\_L & 14.0\tablenotemark{b} & optically thick & 0.03 & 10 & $10^5$ & 2.68 & 1.10 & 0.25 & 2.43 \\
T10\_B0.1\_L & 30.0 & optically thick & 0.1 & 10 & $10^5$ & 6.10 & 2.23 & 0 & 0 \\
T10\_B0.3\_L & 30.0 & optically thick & 0.3 & 10 & $10^5$ & 9.41 & 3.24 & 0 & 0 \\
\enddata
\tablecomments{We cover two different resolutions (Section~\ref{sec:setup:res}) and three different cooling types (Section~\ref{sec:setup:cooling}). The columns show simulation name, duration ($T$), cooling type, cooling model parameters ($\beta_0,\tau_0,c_0$; see definitions in Section~\ref{sec:setup:cooling}), average cooling time of whole disk ($\beta$), average cooling time at $1r_0$ ($\beta_1$), average number of fragments in disk ($n_{\rm frag}$), and \revise{disk-integrated} fragment production rate ($f_{\rm frag}$).}
\tablenotetext{a}{Early termination due to wall clock time.}
\tablenotetext{b}{Early termination because the simulation breaks. For diagnostics, the simulation is cut at a $T$ shortly before evolution becomes unphysical.}
\end{deluxetable*}

\section{Results}\label{sec:results}
\subsection{Overview}
\begin{figure*}
    \centering
    \includegraphics[scale=0.66]{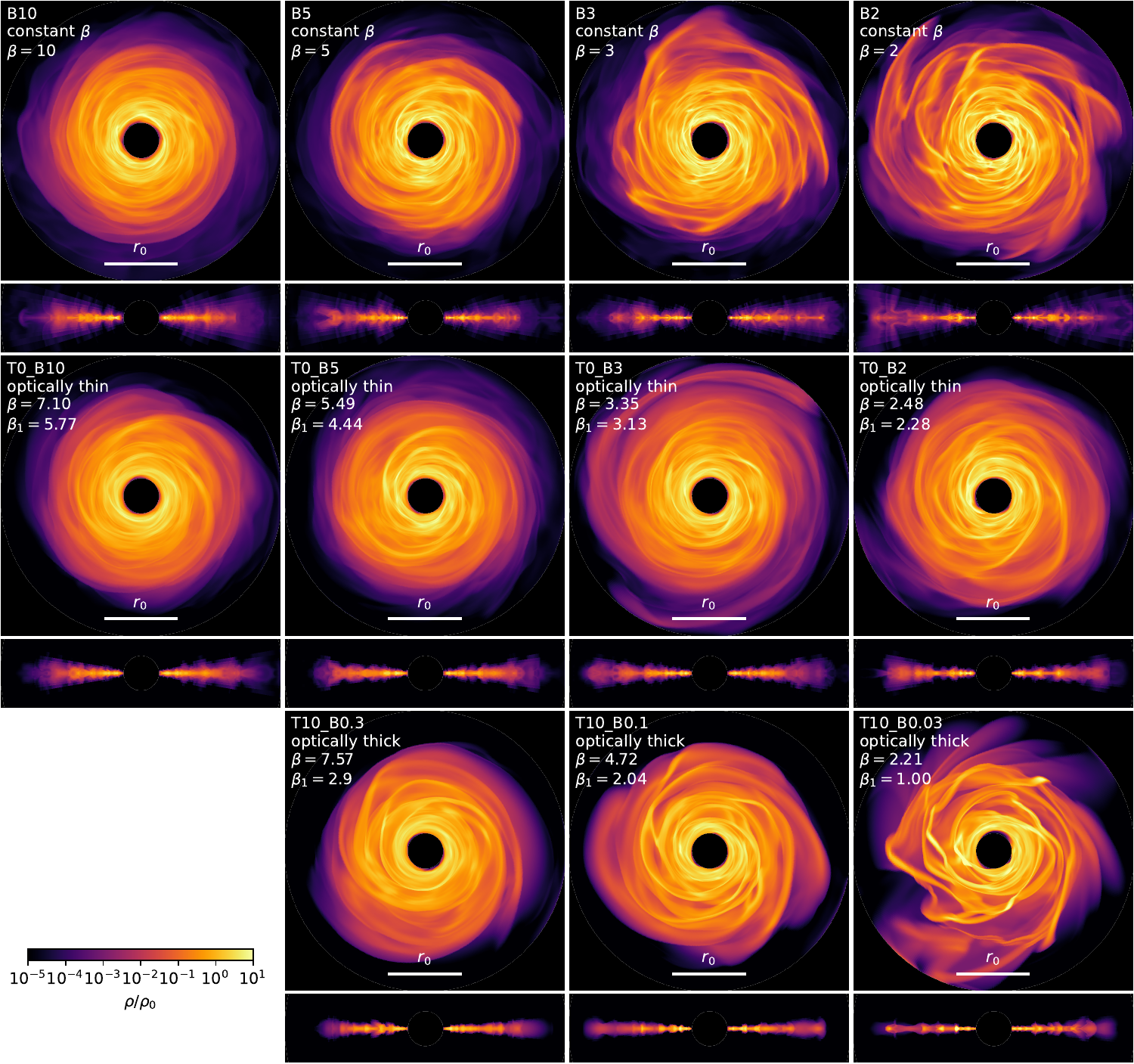}
    \caption{Last snapshots of fiducial resolution simulations. Each row shows a different cooling type. For each run we also label the disk-averaged cooling time $\beta$ and the cooling time at $1r_0$, $\beta_1$; both are averaged across the whole simulation. Disk properties are mainly controlled by the cooling time, and the difference between different cooling types (at similar $\beta$) remains subtle.}
    \label{fig:overview}
\end{figure*}
\begin{figure*}
    \centering
    \includegraphics[width=\textwidth]{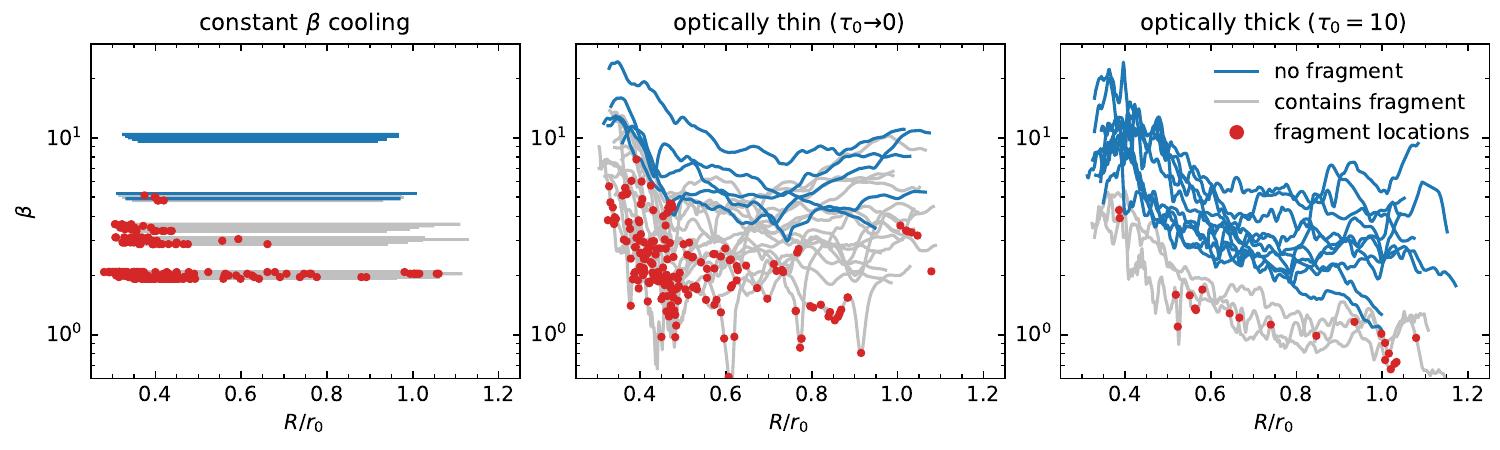}
    \caption{Summary of fragment formation in our simulations. Each line corresponds to a period of $6t_0$ ($\sim$ 1 orbit at $r_0$) in a simulation, and only regions with $Q_{\rm K}\leq2$ are plotted. If no fragment is generated during this period, we color the line blue to mark a non-fragmenting period. Otherwise, the line is colored gray and we mark the locations when fragmentation occurs in red. A fragment that live through multiple snapshots is counted only once, when it first appears. In the left panel, since all lines of the same simulation have the same beta, we add a small displacement to $\beta$ for better viewing. Overall, there is no clear boundary between fragmenting and non-fragmenting regimes due to the stochastic nature of fragmentation. The boundary between red and blue also appears to shift towards larger $\beta$ at small radii, mainly because a given time window corresponds to more orbits at small radii, and that increases the probability of generating at least one fragment.}
    \label{fig:fragmentation_summary}
\end{figure*}
Our simulation parameters are summarized in Table~\ref{tab:runs}. The simulations can be grouped into two resolutions, three cooling types, and a range of different cooling rates ($\beta$) covering the transition from the non-fragmenting regime (up to $\beta\sim 10$) to the violently fragmenting regime. \revise{For radiative simulations, the disk-averaged cooling time $\beta$ is defined by
\begin{equation}
    \beta \equiv \frac{\langle u\Omega_{\rm K}\rangle}{\langle-q\rangle}
\end{equation}
Here $-q$ is the cooling rate per volumn and $u$ is the internal energy per volumn; $\langle\cdot\rangle$ denotes an average in space and time.
Since the cooling rate is generally radius dependent, we also include a measurement of average cooling time at $1r_0$, $\beta_1$.}
Fig.~\ref{fig:overview} presents a gallery of snapshots taken from simulations at our fiducial resolution. We see a clear trend of faster cooling (lower $\beta$) producing higher amplitude perturbations, but beyond that the difference between different cooling types (at similar $\beta$) is subtle.

Table~\ref{tab:runs} and Fig.~\ref{fig:fragmentation_summary} summarize the occurrence of fragmentation in our simulations. \revise{In Table~\ref{tab:runs}, we present the average number of fragments in the disk $n_{\rm frag}$ and the disk-integrated fragment production rate $f_{\rm frag}$, defined as the number of fragment produced in the whole disk per unit time.} In Fig.~\ref{fig:fragmentation_summary}, we plot the radial profiles of cooling rates and label where and whether fragmentation occurs. Two important trends are evident. First, there is no well-defined $\beta$ threshold for fragmentation; one cannot draw a line in $\beta$ that separates fragmenting (red) and non-fragmenting (blue) regimes, and fragmentation persists at high $\beta$. A good example of this is the $\beta=5$ constant-$\beta$ run (B5), which does not show fragmentation for most time and radii but a few fragments are generated in the inner disk. Second, \revise{at a given $\beta$} fragmentation seems to occur more easily in the inner disk \revise{where the same integration time corresponds to a larger number of orbits.}\footnote{\revise{This does not imply that for realistic protostellar disks fragmentation also occurs more easily in the inner disk. The radial variation of $\beta$ is another important factor controlling fragmentation, and for protostellar disks it is generally found that only the outer disk (beyond $50-100$~au) has sufficiently fast cooling to allow fragmentation (\citealt{Rafikov05,WhitworthStamatellos06,Clarke09,RiceArmitage09}; also see \citealt{Xu22}).}} Both observations point to the notion that fragmentation is stochastic, which has been argued by several previous studies \citep{Paardekooper12,HopkinsChristiansen13,Brucy+21}. 

\subsection{Fragment generation rate}\label{sec:results:pfrag}

\begin{figure*}
    \centering
    \includegraphics[width=\textwidth]{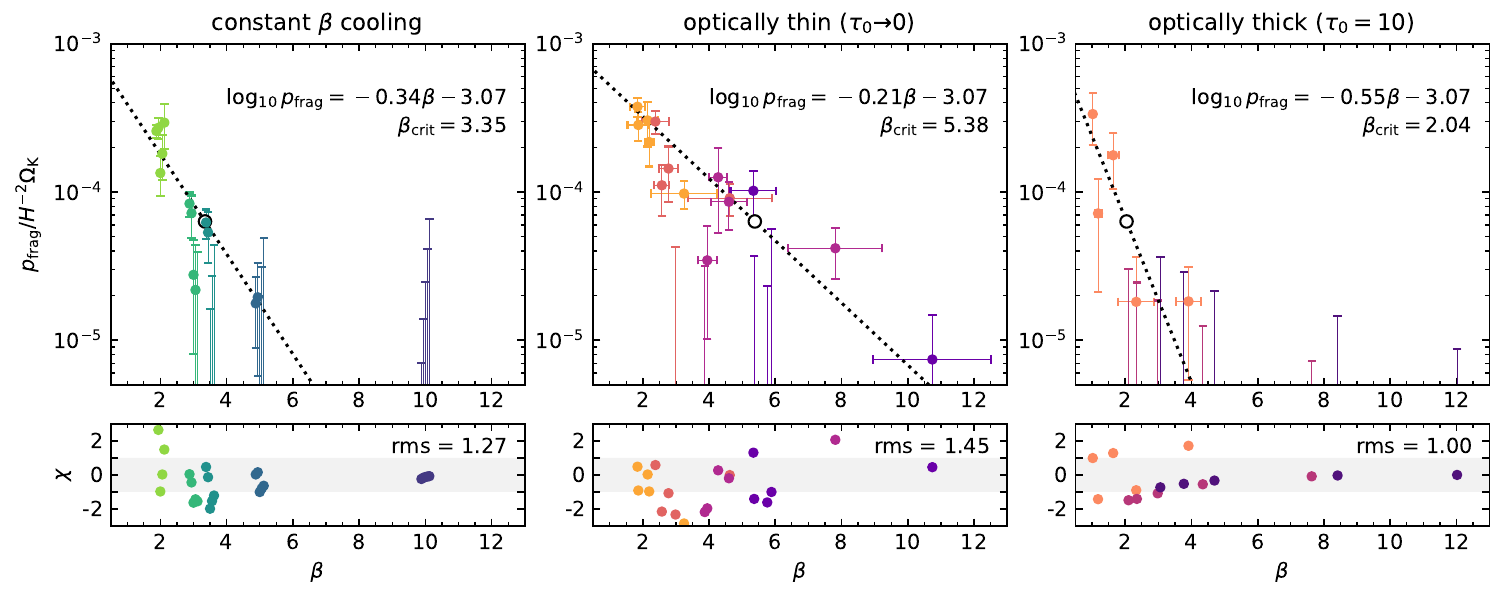}
    \caption{Fragment generation rate $p_{\rm frag}$ as a function of cooling rate. For each simulation we take the range of radii with time-averaged $Q_{\rm K}<2$, and separate it into five rings equally spaced in $\log R$, and each ring corresponds to a data point. The vertical error bar shows the $1\sigma$ statistical uncertainty or upper limit, based on the number of fragments $N_{\rm frag}$ generated in each ring. Different colors encode different simulations. In the left panel, since each simulation has constant $\beta$, we add a small displacement in the horizontal ($\beta$) direction for better viewing. In the center and right panels, the horizontal error bars mark the standard deviation of time-averaged $\beta$ across the radii in the segment. We then fit all data points in a panel to an exponential profile (black dotted lines).
    The bottom panel shows the distribution of the normalized error $\chi \equiv (N_{\rm frag}-N_{\rm pred})/\sqrt{N_{\rm pred}}$, with $N_{\rm pred}$ being the prediction based on the exponential $p_{\rm frag}$ profile. We also plot and label the critical $\beta$ corresponding to $p_{\rm frag}=(2\pi)^{-2}0.05^2 H^{-2}\Omega_{\rm K}$ (white circle), which corresponds to one fragment per orbit per log radius when $H/R=0.05$.}
    \label{fig:fragmentation_probability_together}
\end{figure*}

While the notion of stochastic fragmentation has long existed, there has been no direct measurement of the  fragmentation generation rate per disk area $p_{\rm frag}$ and how it depends on the cooling rate. 
In Fig.~\ref{fig:fragmentation_probability_together}, we produce the first direct measurement of $p_{\rm frag}$. \revise{Within a ring between $R_1$ and $R_2$ and over a time period $T$, $p_{\rm frag}$ can be estimated by
\begin{equation}
    \frac{p_{\rm frag}}{H^{-2}\Omega_{\rm K}} = \frac{N_{\rm frag}}{T\int_{R_1}^{R_2}2\pi RH^{-2}\Omega_{\rm K}{\rm d}R}.
\end{equation}
Here $N_{\rm frag}$ is the number of fragments generated within the ring during $T$. On the right hand side, the scale height $H$ is evaluated using the time-averaged disk profile.}
We evaluate $p_{\rm frag}$ in units of $H^{-2}\Omega_{\rm K}$ because $H$ gives the characteristic length scale of clumps and spirals (\citealt{Cossins+09,Bethune+21}; we also discuss this in paper II).
\revise{In Fig.~\ref{fig:fragmentation_probability_together}, for each simulation we take the range of radii with time-averaged $Q_{\rm K}<2$, separate it into five rings equally spaced in $\log R$, and evaluate $p_{\rm frag}$ on each of the rings for the whole duration of the simulation.}
We plot $p_{\rm frag}$ as a function of the cooling rate $\beta$.
The $\beta$ values are computed from the time and azimuthally averaged cooling rates, and are often not the same as the local $\beta$ values within the clumps/fragments. We choose this definition of $\beta$ because, compared to the local $\beta$ in each clump, it shows less variation during fragmentation and is easier to estimate in disk models.


For each cooling prescription, the fragment generation rate per area $p_{\rm frag}$ (in unit of $H^{-2}\Omega_{\rm K}$) can be fit reasonably well by an exponential profile,
\begin{equation}
    p_{\rm frag} \approx p_0~10^{-f\beta} H^{-2}\Omega_{\rm K}.\label{eq:p_frag}
\end{equation}
Black dotted lines mark the best fit exponential profiles in Fig.~\ref{fig:fragmentation_probability_together}. We require the simulations to share the same intercept $p_0$, since in the limit of fast cooling, all clumps can become fragments and the behavior should be insensitive to cooling type. Relaxing this requirement does not significantly affect the results.
The best-fit exponential profiles are
\begin{equation}
    p_0 = 8.5\times 10^{-4},~~
    f = \begin{cases}
    0.34 & \text{(constant $\beta$)}\\
    0.21 & \text{(optically thin)} \\
    0.55 & \text{(optically thick)}
    \end{cases}.\label{eq:p_frag_params}
\end{equation}

Based on these profiles, we can also define a critical $\beta$ (white circles) corresponding to a fragment generation rate of one fragment per orbit per log radius at a given $H/R$, which we choose to be 0.05.
Roughly speaking, this $\beta_{\rm crit}$ corresponds to the fragmentation ``threshold'', which can be interpreted as the $\beta$ below which generating at least one fragment in the simulation becomes likely. The exact value of this threshold can shift depending on the domain size (in $H^2$) and the duration of the simulation, but such shifts remain small due to the exponential scaling of $p_{\rm frag}$.
We find that
\begin{equation}
    \beta_{\rm crit} = \begin{cases}
    3.35 & \text{(constant $\beta$)}\\
    5.38 & \text{(optically thin)} \\
    2.04 & \text{(optically thick)}
    \end{cases}.\label{eq:beta_crit}
\end{equation}
Our $\beta_{\rm crit}$ for constant $\beta$ runs is consistent with the $\beta\approx3$ fragmentation threshold found by previous studies \citep[e.g.,][]{Gammie01,Deng+17,Baehr+17}.

Comparing across different cooling types, we see that disk thermodynamics does impact the fragment generation rate. Optically thin cooling promotes fragmentation, increasing $\beta_{\rm crit}$ and making $p_{\rm frag}$ decay much slower in $\beta$; this is most clearly shown by the two data points with fragments at $\beta=8-11$ in the middle panel of Fig.~\ref{fig:fragmentation_probability_together}.\footnote{We checked that the fit is not skewed by these two data points. Removing them barely changes the fit, since the data points at lower $\beta$ also suggest a shallow slope.} Meanwhile, optically thick cooling slightly suppresses fragmentation, decreasing $\beta_{\rm crit}$ and making $p_{\rm frag}$ decay more steeply in $\beta$.

The $p_{\rm frag}$ obtained from low-resolution runs are qualitatively similar, but show some quantitative deviation. In Appendix \ref{a:res} we argue that the origin of this difference is mainly the higher numerical viscosity in low-resolution runs, and that the fiducial-resolution runs should be unaffected by numerical viscosity for the range of $\beta$ we surveyed.

\subsection{Initial fragment mass}\label{sec:results:mass}
\begin{figure}
    \centering
    \includegraphics[scale=0.66]{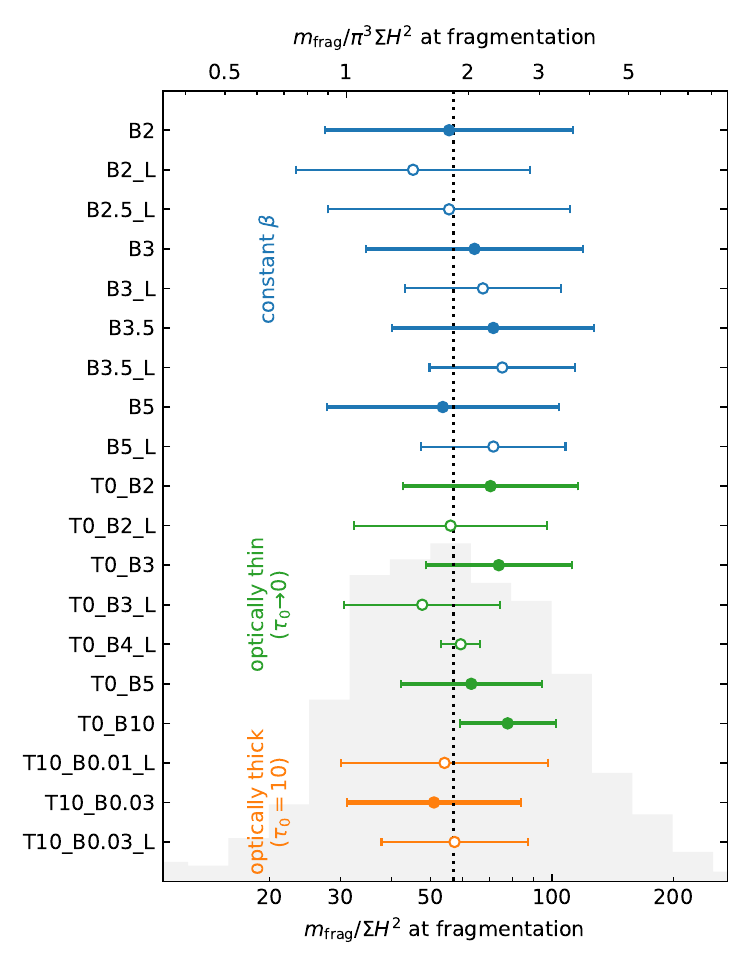}
    \caption{Fragment mass distribution from each simulation. We normalize fragment mass by $\Sigma H^2$ (or $\pi^3\Sigma H^2$, which is a simple analytic estimate) computed from the azimuthally averaged disk profile at the time and radius of fragmentation. For each simulation, we show the logarithmic mean (marker) and $1\sigma$ spread (horizontal error bars). Different colors mark different cooling types, and filled/open markers mark runs with fiducial/low resolution. For each cooling type, the runs are sorted by cooling rate. The combined distribution of all runs are plotted in the gray histogram in the background. All runs show broadly similar $m_{\rm frag}$ distribution.}
    \label{fig:fragment_mass}
\end{figure}

\begin{figure}
    \centering
    \includegraphics[scale=0.66]{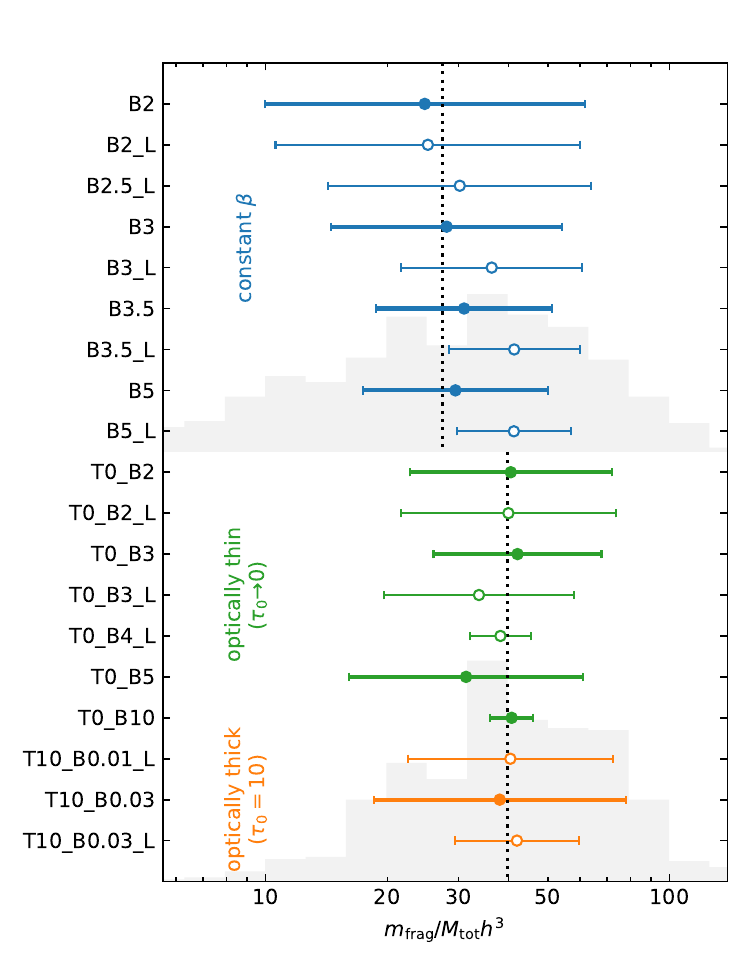}
    \caption{Same as Fig.~\ref{fig:fragment_mass}, but shows fragment mass normalized by $M_{\rm tot}h^3$ where $M_{\rm tot}$ is the total (star+disk) mass and $h=H/R$ is the aspect ratio. Here we use $h$ based on the average disk profile, which is more convenient for predicting fragment masses in semi-analytic calculations. We separately measure the distribution for constant $\beta$ runs and realistic cooling runs, because the two distributions now show nontrivial difference.}
    \label{fig:fragment_mass_2}
\end{figure}

Our simulations also provide a good opportunity to measure the initial mass function of fragments. As we discussed in Section~\ref{sec:fragment_disruption},  fragments tend to be numerically disrupted soon after their formation; this makes it impossible to realistically track long-term fragment evolution, but the maximum bound mass $m_{\rm frag}$ achieved by each fragment in our simulation provides a good estimate for the distribution of initial fragment mass. In our simulations, once a fragment becomes bound, its mass does not increase substantially (usually no more than $10\%$) before it is numerically disrupted.

In Fig.~\ref{fig:fragment_mass} we summarize the $m_{\rm frag}$ distributions from all simulations showing more than one fragment. They show broadly similar distributions, and the combined distribution (gray histogram) is consistent with a log-normal distribution around
\begin{equation}
    m_{\rm frag} \sim 57 \Sigma H^2,
\end{equation}
with $\Sigma$ and $H$ measured using the azimuthally averaged profile at the time and radius of fragmentation. The relatively good agreement across simulations with different cooling rates and cooling types suggests that the initial excitation of clumps may be insensitive to the details of cooling, which is reasonable if the mass contained in a clump is set by the linearly unstable mode that seeds this clump.

We can compare this result with the simple analytic estimate in \citet[Eqs. 49-50]{KratterLodato16}, where the clump material is assumed to come from a circular patch with radius $\lambda/2$ with $\lambda=2\pi H$ being the fastest-growing wavelength at $Q=1$; that yields a mass scale of
\begin{equation}
    M_{\rm clump} = \pi^3 \Sigma H^2 = 31.0 \Sigma H^2.
\end{equation}
\revise{Another simple estimate of the fragment mass is the Jeans mass
\begin{equation}
    M_{\rm Jeans} = \frac{4\pi}{3}\rho\left(\frac{\lambda_{\rm Jeans}}{2}\right)^{3} = 37.7\Sigma H^2.
\end{equation}
Here $\lambda_{\rm Jeans}=\sqrt{\pi c_s^2/G\rho}$ is the Jeans wavelength, and the right hand side is evaluated at $Q=1$ and $\rho=\Sigma/\sqrt{2\pi} H$ (corresponding to the midplane density). Both estimates differ from our results by only a factor of two.
}

Since in a gravitationally unstable disk $Q\sim 1$ and $Q\propto H\Omega_{\rm K}^2/\Sigma$, the typical fragment mass can also be rewritten as $m_{\rm frag}\propto M_{\rm tot}h^3$ where $h=H/R$ is the aspect ratio. This is helpful for providing a straightforward estimate of the fragment mass. In Fig.~\ref{fig:fragment_mass_2} we summarize $m_{\rm frag}$ normalized by $M_{\rm tot}h^3$, and find that
\begin{equation}
    m_{\rm frag} \sim 
    \begin{cases}
      27 M_{\rm tot} h^3 & \text{(constant $\beta$)}\\
      40 M_{\rm tot} h^3 & \text{(radiative cooling)}
    \end{cases}\label{eq:m_frag}
\end{equation}
Here, $h$ is measured from the time- and azimuthally averaged profile; this choice is helpful for semi-analytic disk modeling since the averaged profile, as opposed to the instantaneous profile at fragmentation, is easier to predict.
There is now a nontrivial difference between the prefactor for constant $\beta$ runs and other runs; this is mainly because radiative cooling has a steeper dependence on temperature, which reduces the temperature (and scale height) difference between the clump and the disk.
The spread of $m_{\rm frag}$ is also useful for predicting the outcome of fragmentation. For radiative cooling runs, the distribution of $m_{\rm frag}$ is broadly consistent with log-normal and the $1\sigma$ spread of $\log m_{\rm frag}$ is
\begin{equation}
    \sigma(\log m_{\rm frag}) = 0.57.\label{eq:sig_m_frag}
\end{equation}

\section{The physical picture of stochastic fragmentation}\label{sec:discussion}
\subsection{The basic picture}

\begin{figure*}
    \centering
    \includegraphics[scale=0.66]{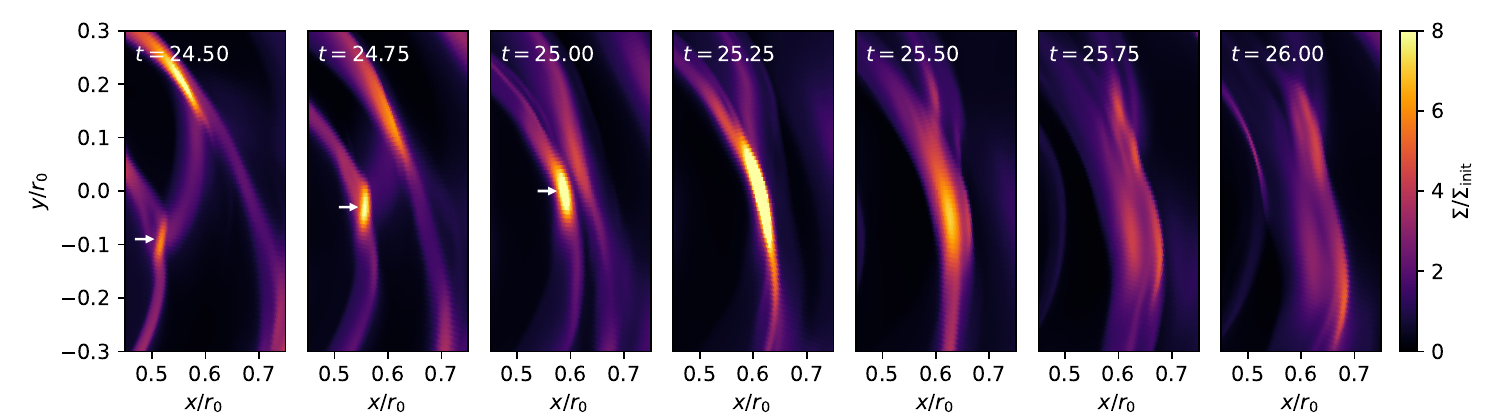}
    \caption{An example of clump disruption by collision, taken from simulation T10\_B0.1. We plot a series of snapshots in a frame corotating at $0.6r_0$. The clump (marked by a white arrow in the first three panels) is initially contracting, but is disrupted after colliding with another clump at $t\approx 25.25$.}
    \label{fig:fragment_collision}
\end{figure*}

The basic picture of stochastic fragmentation can be obtained by looking into the details of how a clump contracts to become a gravitationally bound fragment.
First, consider a clump in isolation without cooling. Self-gravity allows the clump to contract, but not indefinitely; pressure eventually stops the contraction if the adiabatic index satisfies $\gamma>4/3$.
In reality, clumps are neither isolated nor adiabatic, and that introduces two other mechanisms. On the one hand, close encounters with other clumps tend to disrupt the clump. This can happen via direct collision with another clump or the spiral wake/shock it excites, as well as via tidal stripping. In all this cases, clump-clump interaction converts the relative velocity between the clumps to thermal energy and/or velocity dispersion within the clump and that causes disruption. An example of clump disruption via clump-clump collision is given in Fig.~\ref{fig:fragment_collision}. On the other hand, cooling allows contraction to proceed beyond the adiabatic hydrostatic equilibrium. It is the competition between these two mechanisms that determines whether fragmentation can happen: if a clump undergoes sufficient cooling and contraction without encountering any other clump, it can become a bound fragment; otherwise it would be disrupted. An important feature of this picture is that clump-clump encounter is a stochastic event, and that makes fragmentation stochastic. \revise{This idea has been previously discussed by \citet{Paardekooper12} and \citet{YoungClarke15,YoungClarke16}.}

\subsection{Explaining the exponential scaling}
Using the competition between cooling and clump-clump encounter, we can explain the exponential scaling of the fragment generation rate. Let $p_{\rm cl}$ be the rate of clump production per area, and $\Omega_{\rm en}$ the average rate for a given clump to encounter another clump. For simplicity, we assume that clumps evolve independently, so clump-clump encountering becomes a Poisson process. In this case the probability for a given clump to become a bound fragment is simply ${\rm e}^{-\Omega_{\rm en} t_{\rm con}}$ and the fragment generation rate is
\begin{equation}
    p_{\rm frag} = p_{\rm cl} {\rm e}^{-\Omega_{\rm en} t_{\rm con}}.
\end{equation}
Here $t_{\rm con}$ is the contract time, defined as the time it takes for a clump to undergo sufficient contraction (by cooling) so that it becomes too dense to be disrupted by further clump-clump encounters; more precisely, this happens when the energy removal rate by cooling exceeds the typical energy injection rate due to clump-clump encounters. 
We can then define dimensionless encounter and contraction factors $f_{\rm en}$ and $f_{\rm con}$ as
\begin{equation}
    f_{\rm en} \equiv \frac{\Omega_{\rm en}}{\Omega_{\rm K}},~~~
    f_{\rm con} \equiv \frac{t_{\rm con}}{t_{\rm cool}} = \frac{\Omega_{\rm K}t_{\rm con}}{\beta}.
\end{equation}
Here $t_{\rm cool}$ and $\beta$ are the cooling time of the background disk defined using the time- and azimuthally averaged disk profile.
Now the fragment generation rate takes the form
\begin{equation}
    p_{\rm frag} = p_{\rm cl}{\rm e}^{-f_{\rm en}f_{\rm con}\beta}.
\end{equation}
Note the similarity between this formula and Eq.~\eqref{eq:p_frag}, with $p_0 = p_{\rm cl}/H^{-2}\Omega_{\rm K}$ and $f=f_{\rm en}f_{\rm con}/\ln{10}$.

So far we demonstrated that the fragment generation rate can be expressed in a form similar to Eq.~\eqref{eq:p_frag}. We next argue that treating $p_{\rm cl}/H^{-2}\Omega_{\rm K}$ and $f_{\rm en}$ as constants and treating $f_{\rm con}$ as a cooling-type dependent constant are reasonable approximations. First consider $p_{\rm cl}$ and $f_{\rm en}$. For clumps with characteristic size $\lambda$, the surface density of a clump is $n_{\rm cl}\sim 1/\lambda^2$. Assuming that  clumps are all corotating at $\Omega_{\rm K}$ \citep{Cossins+09,Steiman-Cameron+23}, the encounter rate is
\begin{equation}
    \Omega_{\rm en} \sim n_{\rm cl}\int_{-\lambda}^{\lambda}|x\Omega_{\rm K}|{\rm d}x\sim \Omega_{\rm K},~~~f_{\rm en}\equiv \frac{\Omega_{\rm en}}{\Omega_{\rm K}}\sim 1.\label{eq:encounter}
\end{equation}
\revise{An approximately constant $f_{\rm en}$ has also been demonstrated by \citet{YoungClarke16}; they use 2D simulations to measure the distribution of the wait time between encountering two shocks at two different $\beta$ and find $f_{\rm en}\approx 0.2$.}
Meanwhile, since $\Omega_{\rm en}^{-1}$ also serves as the typical clump lifetime, the clump generation rate is
\begin{equation}
    p_{\rm cl} \sim n_{\rm cl}\Omega_{\rm en} \sim \lambda^{-2}\Omega_{\rm K}.
\end{equation}
The clump size $\lambda$ is set by the length scale of the most unstable mode. We expect $\lambda\propto H$ with $\lambda/H$ insensitive to the cooling rate when $\beta\gtrsim 1$, because clump formation, which needs to happen on a timescale $\lesssim\Omega_{\rm en}^{-1}\sim\Omega_{\rm K}^{-1}$, is too quick to be significantly affected by cooling as long as $\beta\gtrsim 1$. The assumption of a nearly universal $\lambda/H$ is also supported by the universal fragment mass observed in Section~\ref{sec:results:mass}. Therefore,
\begin{equation}
    \frac{p_{\rm cl}}{H^{-2}\Omega_{\rm K}} \sim \left(\frac{\lambda}{H}\right)^{-2} \sim {\rm constant}.\label{eq:p_cl}
\end{equation}
Next consider $f_{\rm con}$. For a given cooling type, the cooling timescale of a clump at a given density and temperature profile scales linearly with $\beta$. Thus, the evolution of a contracting clump (unperturbed by clump encounters) as a function of $\beta^{-1} t$ remains similar across different $\beta$.
Therefore, when the level of contraction required for fragmentation is similar, we have 
\begin{equation}
    t_{\rm con}\propto \beta,~~~f_{\rm con}\approx{\rm constant}.
\end{equation}
We note that the arguments above are just rough approximations. In particular, the level of contraction required for fragmentation may have some slight dependence on $\beta$, since for faster cooling the clump may not need to contract as much to become insensitive to energy injection through encounters. Still, we may regard the exponential profile as a reasonable approximation that captures the leading-order $\beta$ dependence of $\log p_{\rm frag}$.

\subsection{The effect of radiative cooling}\label{sec:discussion:radiative}

Using the theoretical framework above, we can also explain the dependence of $p_{\rm frag}$ on cooling type. This mainly comes from the variation of $f_{\rm con}$. To achieve a similar level of contraction, the time required is directly proportional to the local cooling timescale at the clump, $\beta_{\rm cl}$. In other words, comparing across different cooling types, we expect
\begin{equation}
    f_{\rm con} \propto \frac{\beta_{\rm cl}}{\beta}.
\end{equation}
One caveat is that $\beta_{\rm cl}$ is generally not a single value as it can vary during contraction, and $\beta_{\rm cl}$ above should be averaged across the contraction process.

For constant $\beta$ cooling, $\beta_{\rm cl}/\beta=1$ by definition. But for radiative cooling, the clump cooling time $\beta_{\rm cl}$ deviates from the disk cooling time $\beta$.
In the limit of optically thin/thick cooling, the cooling rate per disk area scales as (see Eq.~\ref{eq:cooling_analytic})
\begin{equation}
    \Lambda \propto \begin{cases}
T^4\tau &\text{(optically thin)}\\
T^4\tau^{-1} &\text{(optically thick)}
\end{cases},
\end{equation}
and the cooling time scales as
\revise{
\begin{equation}
    t_{\rm cool} \propto \frac{\Sigma T}{\Lambda} \propto \begin{cases}
T^{-3} &\text{(optically thin)}\\
T^{-3}\tau^2 &\text{(optically thick)}
\end{cases}.
\end{equation}
}
For optically thin cooling, the hotter clump cools faster and $\beta_{\rm cl}/\beta<1$, making the $p_{\rm frag}$ profile shallower.
For optically thick cooling, the increased optical depth reduces cooling. Assuming that the clump undergoes similar levels of contraction along all directions, the temperature and optical depth scales with the contraction factor $\xi$ as $T\propto \xi^{-1}$ and $\tau\propto \xi^{-2}$, so $t_{\rm cool}\propto \xi^{-1}$ increases as contraction proceeds. Therefore, $\beta_{\rm cl}/\beta>1$ and the $p_{\rm frag}$ profile becomes steeper. This explains how optically thin/thick cooling promotes/suppresses fragmentation.

\subsection{The potential importance of opacity profile}\label{sec:discussion:opacity}

In our simulations we only consider a constant opacity, but in protostellar disks the opacity mainly comes from dust grains and is temperature dependent. Below $\sim 150$~K, dust opacity increases with temperature (see Fig.~1 of \citealt{Xu22}), and it further increases/decreases clump cooling rate in the optically thin/thick regime. That causes both effects mentioned above to be more pronounced.

Considering a temperature-dependent opacity also opens up an interesting regime not covered by our theory. When the temperature dependence of opacity is sufficiently steep \revise{and the disk is optically thick, a clump} 
may never become insensitive to clump-clump encounters. This happens when a clump is never able to remove energy (via cooling) fast enough to counter energy injection via clump-clump encounters at a typical rate. Even for a constant opacity, the optically-thick energy removal rate per mass ($\propto T^4\tau^{-2}$) is only constant during contraction, and it further decreases if we have an opacity increasing with temperature.
The critical value of ${\rm d}\log\kappa/{\rm d}\log T$ required to enter this regime of insufficient clump cooling will depend on the details of clump-clump encounter, especially how the energy injection rate by encounter scales with the level of contraction. Within this regime, the theory of fragmentation discussed above breaks down; the disk either never fragments, or fragments through a qualitatively different mechanism \revise{(e.g., via a different source of cooling such as convection)}. 
\revise{Some previous studies have observed fragmentation while adopting a more realistic dust opacity that captures the positive ${\rm d}\log\kappa/{\rm d}\log T$ at low temperature \citep[e.g.,][]{MercerStamatellos17,Hall+17,Cadman+21}, but it remains unclear whether fragmentation occurs at optically thick portions of the disk in these simulations. The dyanmics of this regime, and whether it is relevant for realistic protostellar disks, will be left for future studies.}



\section{Fragmentation in protostellar disks: brown dwarfs or planets?}\label{sec:application}

\begin{figure}
    \centering
    \includegraphics[scale=0.66]{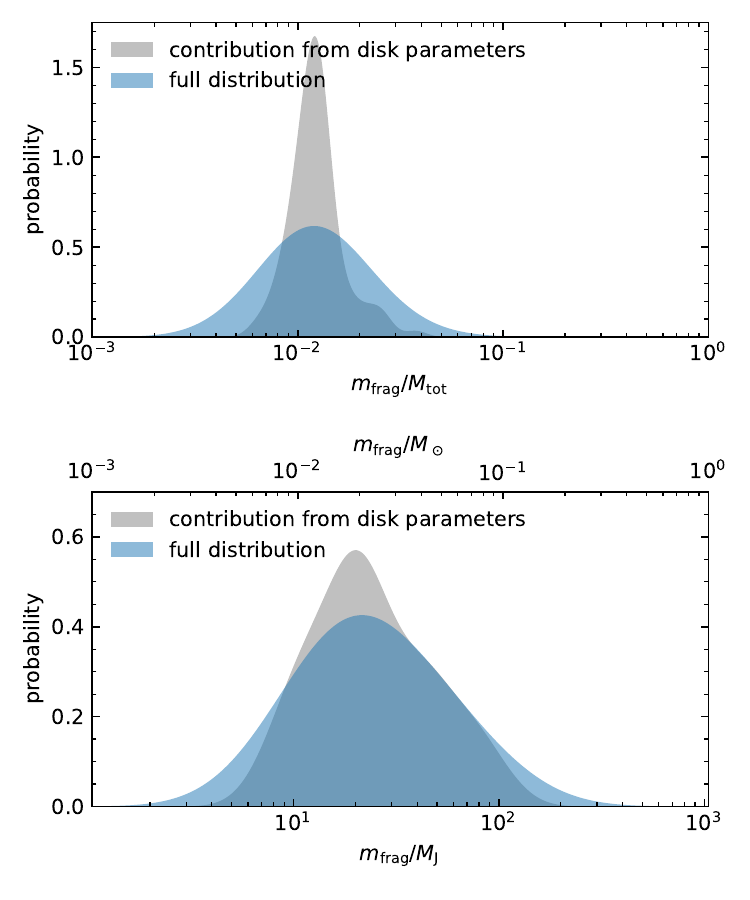}
    \caption{Estimated fragment mass distribution (blue shades) based on $\beta_{\rm crit}$ and $m_{\rm frag}$ from Section~\ref{sec:results} and disk profiles estimated for Orion Class 0/I disks in \citet{Xu22}. We plot the fragment-total mass ratio (top panel) and the fragment mass in Jupiter and solar mass (bottom). For each panel, we first plot the distribution of typical fragment mass (Eq.~\ref{eq:m_frag}) due to variation of disk parameters (grey shades) and then convolve it with the intrinsic spread of $m_{\rm frag}$ (Eq.~\ref{eq:sig_m_frag}) to obtain the full distribution (blue shades).}
    \label{fig:m_frag_orion}
\end{figure}

\subsection{Initial fragment mass}\label{sec:application:frag_mass}

The initial fragment mass ($m_{\rm frag}$) distribution serves as a rough estimate of the mass distribution of companions formed via disk fragmentation, under the caveat that the fragment may gain or lose mass during subsequent evolution as discussed in the next subsection. Below we predict the distribution of $m_{\rm frag}$ in fragmenting protostellar disks and compare that against observational constraints.

In Section \ref{sec:results:mass}, we find the fragment mass distribution can be approximated by a log-normal distribution centered at (Eq. \ref{eq:m_frag})
\begin{align}
    m_{\rm frag} &\sim 40 M_{\rm tot}h^3\\
    &\sim 19 \left(\frac{M_{\rm tot}}{1~{\rm M}_\odot}\right)^{-1/2}\left(\frac{r}{50~{\rm au}}\right)^{3/2} \left(\frac{T}{30~{\rm K}}\right)^{3/2} {\rm M}_{\rm J},\nonumber
\end{align}
with a $1\sigma$ logarithmic spread of 0.57 ($\sim$ a factor of 2; Eq \ref{eq:sig_m_frag}).
To predict the typical value and overall distribution of $m_{\rm frag}$, we also need to know the distribution of $M_{\rm tot}$, $r$, and $T$ at fragmentation. For this, we adopt the disk population in \citet{Xu22}, which is based on a semi-analytic model of gravitationally self-regulated disks. The parameters of this model are fitted using multi-wavelength dust-continuum maps of Class 0/I disks in the Orion molecular clouds. For each disk, we use $\beta_{\rm crit}$ in Section \ref{sec:results:pfrag} to estimate whether it fragments; if it fragments, we pick the smallest radius with $\beta<\beta_{\rm crit}$ and estimate $m_{\rm frag}$ based on the disk properties there. The resulting $m_{\rm frag}$ distribution is summarized in Fig. \ref{fig:m_frag_orion}.

Our predicted $m_{\rm frag}$ distribution is centered at ${\sim}20{\rm M}_{\rm J}$. The overall distribution is relatively wide, due to the spread in stellar and disk properties as well as the intrinsic spread of $m_{\rm frag}$. If $m_{\rm frag}$ directly translates to the companion mass, this result would suggest that fragmentation typically forms brown dwarfs, \revise{in agreement with previous analytic and semi-analytic estimates \citep[e.g.,][]{WhitworthStamatellos06,NeroBjorkman09,Kratter+10,Rice+15}}. Meanwhile, forming massive planets is still quite possible given the width of the distribution.

We then compare this result with observation. The IMF in star-forming regions show secondary peaks at ${\sim}20$--$25{\rm M}_{\rm J}$ \citep{Drass+16,Suarez+19}, consistent with our prediction. Directly imaged companions with separations 10--100~au around low-mass (FGKM) stars also show a similar peak at ${\sim}20{\rm M}_{\rm J}$ \citep{Nielsen+19,Vigan+21}. However, these studies also find that companions around massive (BA) stars are more often ${\sim}10{\rm M}_{\rm J}$ planets. These companions are systematically less massive than those we predict to form via disk fragmentation, and may be an indication that companions around massive stars form by other means, e.g., via core accretion. \revise{The whole population of observed massive sub-stellar companions also show two distinct sub-populations separated at $4-10M_{\rm J}$ \citep{Santos+17,Schlaufman18}; the two sub-populations are broadly consistent with formation by core accretion and gravitational fragmentation, respectively. One caveat is that detection biases may contribute to such trend when combining results from different detection methods.}
Another important constraint comes from AB~Aur, which is so far the only system that contains a candidate protoplanet  embedded in a disk showing clear evidence of GI \citep{Currie+22}. The estimated planet mass is relatively low and marginally consistent with our prediction. The observationally estimated $m_{\rm frag}/M_{\rm tot}$ (this provides a more stringent constraint than $m_{\rm frag}$ alone because it has a narrower distribution; see Fig.~\ref{fig:m_frag_orion}) is $10{\rm M}_{\rm J}/2.4{\rm M}_\odot=4\times 10^{-3}$ and lies at $1.7\sigma$ of our predicted distribution.

\subsection{Subsequent evolution}
Our simulations cover only fragment formation but not subsequent fragment evolution. After formation, the fragment may further increase mass via accretion or decrease mass via tidal stripping \citep[e.g.,][]{Nayakshin10}; it may also migrate in the disk, and may eventually be accreted by the protostar \citep[e.g.,][]{Zhu+12}, which also serves as a possible cause of FU Ori events \citep[e.g.,][]{Vorobyov+21}. How much will these processes affect fragment evolution and survival? Borrowing insights from planet-disk interaction, one might argue that the fragment, with its high initial mass, would open a gap in the disk and that reduces the mass and angular momentum exchange between the planet and the disk. More quantitatively, the gap-opening criterion is \citep[see Section~2.4 of][]{Paardekooper+23}
\begin{equation}
    q \gtrsim 5 h^{5/2} \alpha^{1/2},
\end{equation}
where $q$ is the planet-star mass ratio. Taking steller mass $\approx M_{\rm tot}$ and $\alpha\sim 1$, we have
\begin{equation}
    \frac{m_{\rm frag}}{m_{\rm gap}} \sim 1.8\left(\frac{h}{0.05}\right)^{1/2}.
\end{equation}
Here $m_{\rm gap}$ is the threshold mass for gap opening. This suggests that a typical fragment may open a gap in the disk. We caution that the above threshold is calibrated with viscous disks with $\alpha\ll 1$, and it is not obvious whether it remains applicable in the presence of gravitortubulence at order-unity $\alpha$. For example, gap-opening is not common in the 3D simulations of fragmentation and subsequent fragment evolution by \citet{Boss23}, although this may be related to the finite integration time ($\sim 1000$yr, or $\sim 30$ orbits).
Even without gap opening, the turbulent disk environment and the presence of other fragments cause the evolution of a fragment to deviate from the traditional picture of migration and accretion in a viscous disk; the strong perturbations can suppress migration, and may introduce enough perturbation to eject the fragment or produce sufficiently high eccentricity and/or inclination to substantially reduce accretion \citep{Boss23,Wu+24}.
Finally, the fragment may also suppress GI by driving angular-momentum transport and by heating the disk with the luminosity of the contracting and accreting protoplanet/proto-brown-dwarf \citep{MercerStamatellos17}; this both limits further fragment production and promotes gap opening.
All these possibilities need to be tested with future 3D simulations with longer integration time and better sink prescriptions for fragments.

\subsection{\revise{Comparison with previous 3D simulations}}

\revise{Among previous studies measuring fragment mass distribution from global 3D simulations, a wide range of results have been found. In hydrodynamic simulations (with and without radiation), the typical fragment mass ranges from $\sim M_{\rm J}$ to few $100 M_{\rm J}$ \citep{Stamatellos+07,StamatellosWhitworth09,Hall+17,MercerStamatellos17,Boss21,BossKanodia23}; our estimate of $\sim 20M_{\rm J}$ also lies within this range.

What causes the large scatter across different studies? One possible origin is the aspect ratio $h$ (or, equivalently, the local disk-to-star ratio $R^2\Sigma/M_{\rm tot}$; the two are directly proportional when $Q\sim 1$). As we discussed in Section \ref{sec:results:mass}, the typical fragment mass is $m_{\rm frag}\propto M_{\rm tot}h^3$, and small differences in $h$ can translate to much larger differences in $m_{\rm frag}$. Unfortunately, we cannot quantitatively determine whether this explains most of the scatter in $m_{\rm frag}$ because previous 3D simulations seldom include measurements of $h$ or $M_{\rm frag}/M_{\rm tot}h^3$.
We also note that a key difference between our estimate in Section \ref{sec:application:frag_mass} and previous measurements from 3D simulations is that previous studies directly adopt the fragment masses in simulations, whereas we use simulations to calibrate the dimensionless parameter $m_{\rm frag}/M_{\rm tot}h^3$ and then model observational data to calibrate $h$ at the location of fragmentation. Our approach overcomes the limitation that the aspect ratio of simulated disks are heavily affected by artificial initial conditions, whose parameters may or may not accurately reflect realistic protostellar disks.

The inclusion of additional physical processes may also affect the fragment mass. \citet{Deng+21} and \citet{Kubli+23} find that magnetic field and non-ideal MHD effects may reduce fragment mass down to $\sim 0.01 M_{\rm J}$, with the caveat that the resistivity adopted in these studies may or may not be realistic for protostellar disks.

Finally, fragmentation may also be affected by numerical issues such as insufficient resolution and excessive angular-momentum dissipation (due to artificial viscosity in SPH codes or orbital advection in grid-based codes; see \citealt{Rice+14} and \citealt{Deng+17}). These issues can be detected with resolution studies, as their effects are generally resolution dependent. The two different resolutions in our study yield similar fragment mass distributions. Meanwhile, most previous fragment mass measurements do not include systematic resolution studies, leaving some uncertainties on whether numerical issues affect their results.}
%

\section{Conclusion}\label{sec:conclusion}
In this paper, we use 3D global hydrodynamic and radiation hydrodynamic simulations to investigate fragmentation in gravitationally unstable accretion disks. Below we summarize our main findings.
\begin{itemize}
\item Fragmentation is stochastic, and the fragment generation rate shows an exponential scaling in $\beta$ (Fig.~\ref{fig:fragmentation_probability_together} and Eqs.~\ref{eq:p_frag}--\ref{eq:p_frag_params}). The exponential scaling can be explained as the probability of a clump to avoid close encounter with other clumps until it undergoes sufficient cooling and contraction (Section~\ref{sec:discussion}). The fragment generation rate shows some dependence on cooling type: compared to constant $\beta$ cooling, radiative cooling in the optically thin/thick regime shows a shallower/steeper dependence on $\beta$ (i.e., easier/harder to fragment). This can be explained by the decrease/increase of cooling timescale as the clump contracts (Section~\ref{sec:discussion:radiative}).
\item Due to this exponential dependence, the threshold $\beta$ value corresponding to generating $\mathcal O(1)$ fragment per orbit is only weakly affected by cooling type; it is ${\approx}3$ for constant $\beta$, and becomes ${\approx}5$ and ${\approx} 2$ for optically thin and thick radiative cooling, respectively (Eq.~\ref{eq:beta_crit}).
\item The initial fragment mass is remarkably uniform across different cooling rate and cooling types. For the purpose of modeling realistic (radiatively cooled) disks, the typical initial fragment mass is $40 M_{\rm tot}h^3$ with a spread of $\sigma(\log m_{\rm frag})=0.57$ (Section~\ref{sec:results:mass}).
\item In the context of protostellar disks, our typical initial fragment mass is ${\sim} 20M_{\rm J}$ and lies in the brown dwarf regime. However, given the width of the initial fragment mass distribution and the diversity of disks, forming massive gas giants is still possible (Fig.~\ref{fig:m_frag_orion}).
\end{itemize}

\section*{}
\noindent
We thank the anonymous referee for providing constructive and insightful comments.
W.X. thanks Philip Armitage, Hongping Deng, Ruobing Dong, and Giuseppe Lodato for insightful discussions.
Simulations in this work are performed with computational resources at the Flatiron Institute.

\appendix
\twocolumngrid

\section{Analytic cooling rate}\label{a:cooling_analytic}

\begin{figure}
    \centering
    \includegraphics[scale=0.66]{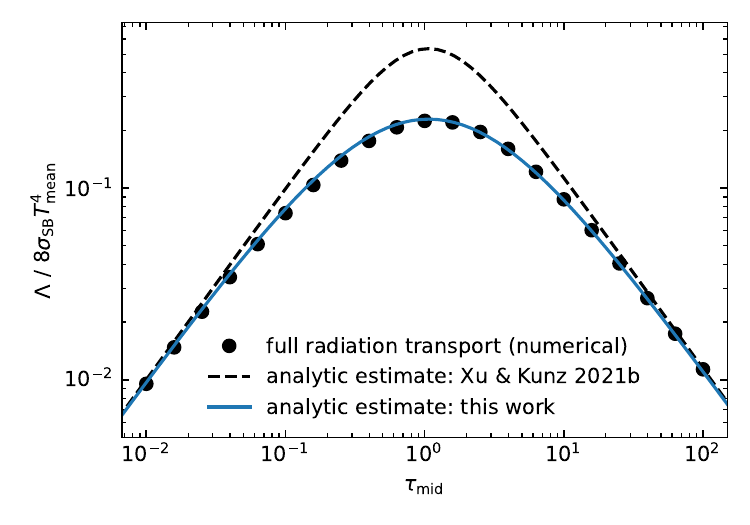}
    \caption{The cooling rate in a geometrically thin, uniformly heated disk with constant opacity. We compare between exact numerical solution (black dots) and analytic formulae from Eq.~(10) in \citet[][black dashed line]{XK21b} and Eq.~\eqref{eq:cooling_analytic} from this work (blue line). Through the introduction of a fudge factor, we significantly improve the accuracy of the interpolation between the optically thin and thick limits.}
    \label{fig:analytic_cooling}
\end{figure}

The analytic cooling rate in Eq.~\eqref{eq:cooling_analytic} is used to approximate the cooling rate in a disk with uniform opacity (only absorption, no scattering) and in thermal equilibrium with a uniform heating rate per mass. In the limit of a geometrically thin disk, the problem is 1D since the disk is effectively infinite in horizontal directions. Eq.~\eqref{eq:cooling_analytic} is based on Eq.~(10) from \citet{XK21b}, which interpolates between the exact analytic results in the limits of $\tau_{\rm mid}\to0$ and $\tau_{\rm mid}\to\infty$. We improve this cooling rate formula by smoothing the interpolation with a fudge factor $p$ in the $\tau_{\rm mid}\sim 1$ regime,
\begin{equation}
    \Lambda_{\rm a}(\tau_{\rm mid}, T_{\rm mean}) \equiv \frac{8\sigma_{\rm SB}\tau_{\rm mid}T_{\rm mean}^4}{[1+(0.875\tau_{\rm mid}^2)^p]^{1/p}}.
\end{equation}
The formula in \citet{XK21b} corresponds to $p=1$. Here we fit $p$ with numerical solution of this 1D radiation transport problem. With the best-fitting $p=0.45$, it approximates the cooling rate for all optical depths with error of at most a few percent (Fig.~\ref{fig:analytic_cooling}).

This formula can also be easily modified to account for the presence of external heating if we still assume a gray opacity without scattering. For external heating with an incoming flux per disk area of $\sigma_{\rm SB}T_{\rm ext}^4$ (this may and may not be isotropic), we simply need to replace $T_{\rm mean}^4$ by $T_{\rm mean}^4-T_{\rm ext}^4$ to get the net cooling rate, defined as cooling minus external heating (so net cooling balances internal heating in thermal equilibrium). This is because in the absence of scattering a solution to the radiation transport equation remains valid under any offset that is uniform in space and constant in time $I(\mu,z,t)\to I(\mu,z,t)+\Delta I(\mu)$, where $I(\mu,z,t)$ is the intensity and $\Delta I(\mu)$ is the offset. External heating can be captured by adding a $\Delta I(\mu)$ following the incoming irradiation at $z=\pm\infty$.

\section{Unphysical disruption of fragments}\label{a:disruption}

A fragment that would physically collapse could become disrupted in our numerical simulation due to finite resolution. Once the collapse proceeds to grid scale and the density contrast between adjacent cells becomes large, the finite resolution can lead to unphysical energy injection which ultimately disrupts the clump, especially when the clump is being advected quickly through the grid. Below we demonstrate this effect with a toy problem.

Consider a periodic box with $x,y,z\in[-2,2]$ and $32^3$ cells, initialized with the following profile:
\begin{align}
    &\rho = 1 + \exp[-(x^2+y^2+z^2)/2],\\
    &p = 1,\\
    &v_x, v_z = 0,\\
    &v_y = 10.  
\end{align}
We include self-gravity with $G=1$ and cooling with $t_{\rm cool}=1$. This toy problem represents a collapsing clump in a flow that quickly advects through the grid.

\begin{figure}
    \centering
    \includegraphics[scale=0.66]{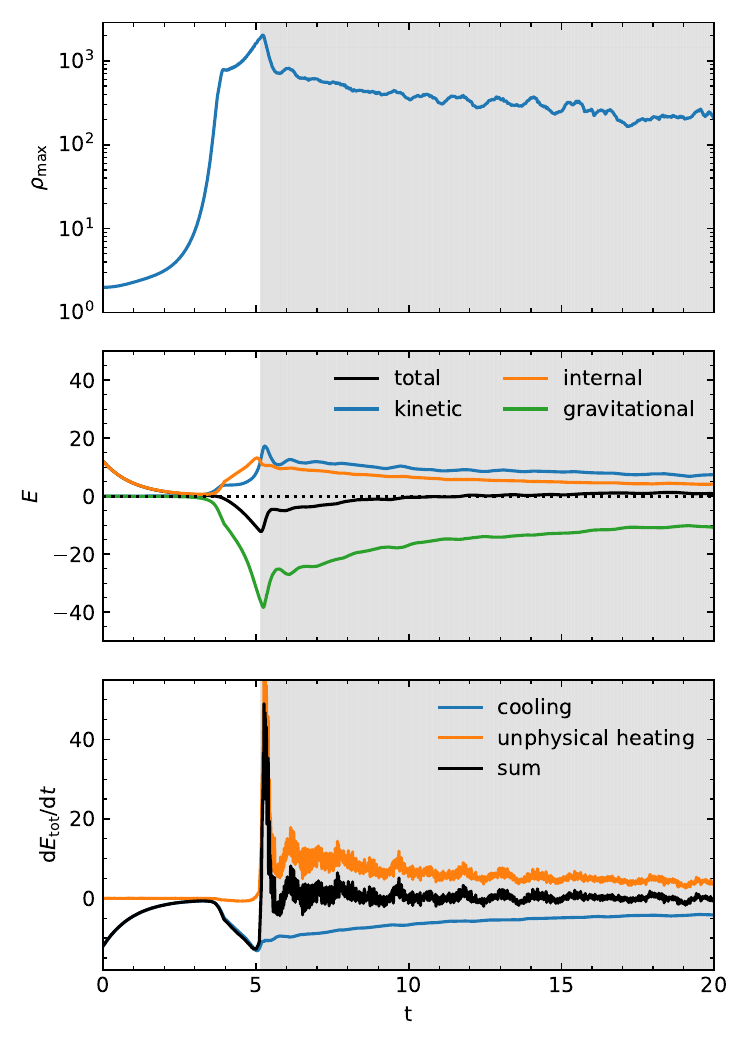}
    \caption{Density and energy evolution of our toy model of numerical fragment disruption. The energy in the middle panel is evaluated in the frame moving at the initial $v_y$. The gray shaded regions mark when the unphysical heating is $\geq 50\%$ of the physical cooling. Once collapse proceeds to grid scale, unphysical heating injects energy to unbind the contracting clump.}
    \label{fig:v10}
\end{figure}

\begin{figure}
    \centering
    \includegraphics[scale=0.66]{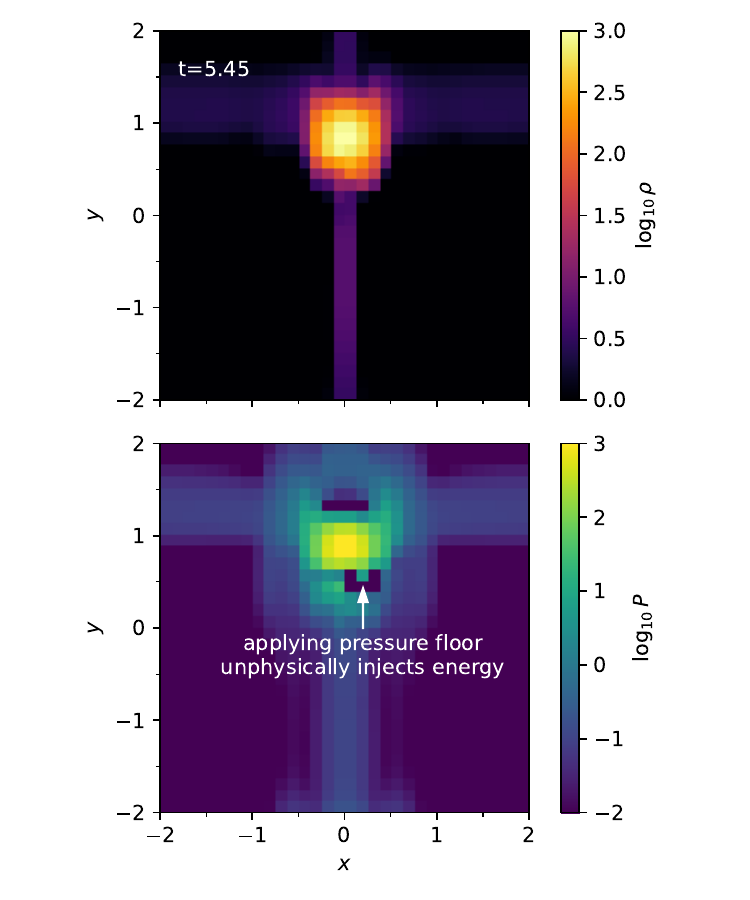}
    \caption{A $z=0$ slice in our toy model of numerical fragment disruption, at a time when strong unphysical heating occurs.}
    \label{fig:pressure}
\end{figure}

The time evolution of density and energy is shown in Fig.~\ref{fig:v10}. During the beginning of the collapse, the clump remains relatively well resolved and collapse proceeds normally. But once the fragment becomes unresolved at $t\approx 5$, the collapse stops and density begins to decrease. Comparing ${\rm d}E/{\rm d}t$ measured from the simulation with the cooling (which is the only physical cause of total energy change in the box) shows that unphysical energy injection is present when the clump becomes unresolved. This can unbound the clump by increasing the total energy. In this toy model the clump eventually settles into a state where unphysical heating balances physical cooling, but in our gravitoturbulent disk simulations such energy injection tends to disrupt the clump once it no longer has sufficiently negative energy to resist turbulent and tidal perturbations.

How does this unphysical energy injection happen? Once the clump is sufficiently collapsed, the large density contrast between cells means that for the lower-density cells around the pressure maximum, the fluxes and flux divergences may be large compared to the amount of mass, momentum, and energy in the cell, and some cells can be left with negative thermal energy after applying the flux divergences. But this negative thermal energy will be removed when we apply the pressure floor, causing unphysical heating (Fig.~\ref{fig:pressure}).

As a side note, although the current self-gravity source term in \texttt{Athena++} does not exactly conserve energy, it is not the dominant source of error. We tested the exactly energy-conserving prescription in \citet{Mullen+21} on the toy problem, and it produces similar results.

This behavior has not been reported in previous simulations (including \texttt{Athena++} shearing box simulations such as \citealt{Chen+23}) for two main reasons. First, some simulations keep the fragment well-resolved by increasing resolution or artificially limiting the level of collapse. For example, in setups with fixed mass resolution (e.g., SPH), the clump remains well-resolved. Simulations that include some smoothing length for self-gravity such as 2D thin-disk simulations suppresses collapse beyond a certain scale and can leave the fragment at a finite size. These treatments have their limitation though. With adaptive spatial resolution, it becomes numerically unfeasible to evolve the disk for an extended period of time beyond the first fragmentation event. With smoothed self-gravity, keeping the fragment at a finite size (often comparable to $H$) may artificially increase its mass exchange with the disk.

Second, once the fragment becomes unresolved, what kind of unphysical behavior it exhibits can vary depending on the details of the numerical setup and the physical properties of the clump. In particular, whether the clump quickly advects through the grid can affect the behavior. In our toy problem, when the clump advects quickly, we tend to see unphysical heating and fragment disruption. But when the clump advects slowly, we tend to see the clump collapsing to grid scale and breaks the run, likely by leaving a cell with large momentum but near-zero (or negative) density; the application of the small density floor (while conserving momentum) then produces enough kinetic energy to break the simulation by producing overflow. The regime of slow advection better resembles a shearing box setup. The key difference between slow and fast advection is that for fast advection, advection smears out the density maximum more, and the region around the maximum has higher density and the simulation is less likely to break.

\begin{figure*}
    \centering
    \includegraphics[width=\textwidth]{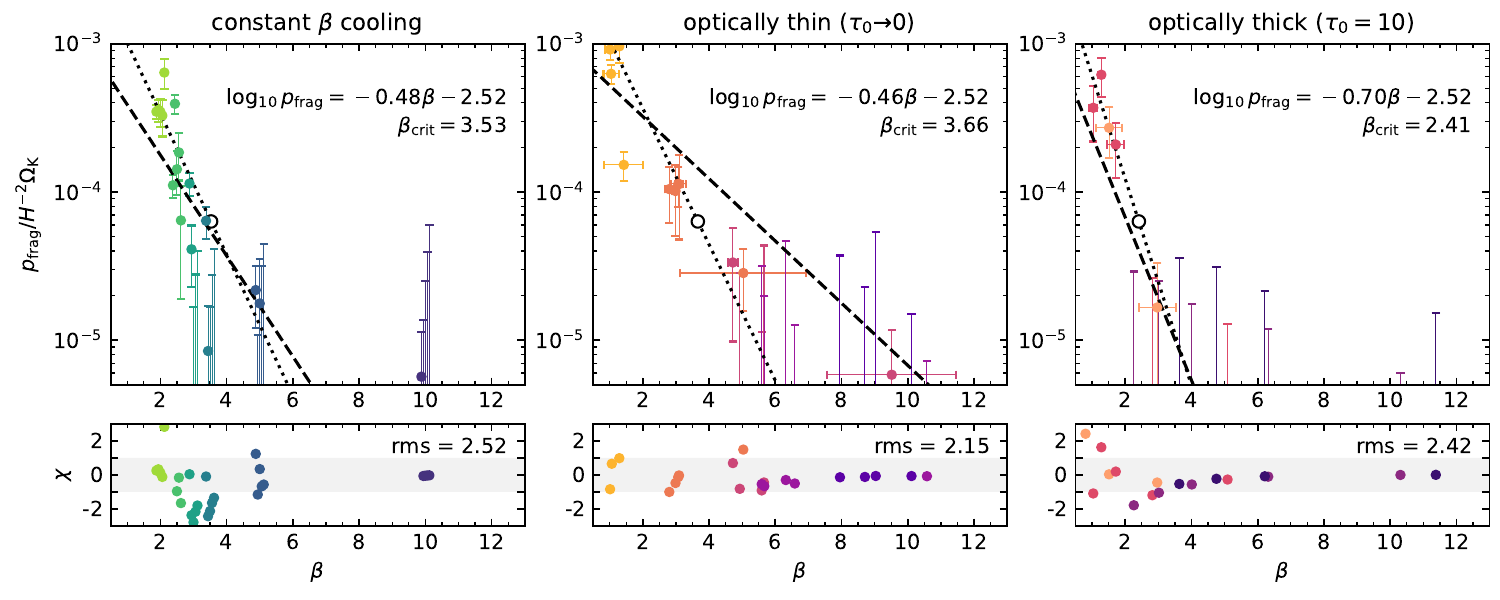}
    \caption{Similar to Fig.~\ref{fig:fragmentation_probability_together}, but showing results from low-resolution runs. The exponential fits in Fig.~\ref{fig:fragmentation_probability_together} are shown in black dashed lines for comparison. The low-resolution runs show steeper $p_{\rm frag}$ profiles, especially for the optically thin runs. Such difference can be attributed to a higher level of numerical viscosity.}
    \label{fig:fragmentation_probability_lowres_together}
\end{figure*}

\section{Resolution dependence of fragmentation probability}\label{a:res}

Among results presented in this paper, the only case where low and fiducial-resolution runs show a nontrivial difference is the fragmentation probability.
As demonstrated in Fig.~\ref{fig:fragmentation_probability_lowres_together}, the low-resolution runs generally show higher/lower fragment generation rate at short/long cooling timescale. This produces steeper $p_{\rm frag}(\beta)$ profiles, which is especially significant for optically thin runs because the fiducial resolution optically thin runs show nontrivial $p_{\rm frag}(\beta)$ up to $\beta\sim 10$ but the low-resolution runs rarely fragment in that regime.

The difference can be interpreted as a result of numerical viscosity due to finite resolution. The role of numerical viscosity is twofold. On the one hand, by forcing numerical mass diffusion near grid scale, it suppress the contraction of clumps. On the other hand, it converts kinetic energy into thermal energy, which could promote fragmentation in the regime of rapid cooling because the thermal energy can be quickly removed. Together, these effects explain why the low-resolution runs show higher/lower fragmentation rate at fast/slow cooling.

Does numerical viscosity also nontrivially affect the results of our fiducial-resolution runs? To answer this question, we can estimate the $\beta$ value beyond which numerical viscosity significantly suppresses fragmentation, denoted by $\beta_{\rm visc}$. This happens, roughly speaking, when $\Omega_{\rm K}^{-1}\beta$, which sets the minimal timescale scale of contraction, is above the timescale for numerical viscosity to diffuse a typical clump, $t_{\rm visc}$. At a given physical scale, $t_{\rm visc}$ scales as $\Delta^{-3}$ where $\Delta$ is the cell size (see tests in Appendix B of \citealt{XuThesis}). Form low resolution to fiducial resolution, the midplane resolution doubles, and that increases $\beta_{\rm visc}$ by a factor of 8. Meanwhile, judging by the fact that low-resolution runs still fragment for $\beta\sim 3$ and have fragment generation rate similar to the fiducial-resolution runs, we expect $\beta_{\rm visc}\gtrsim 3$ for low resolution and thus $\beta_{\rm visc}\gtrsim 24$ for fiducial-resolution runs. This suggests that our estimated fragmentation rates in Section~\ref{sec:results:pfrag}, which are based on results for $\beta\lesssim 10$, should remain unaffected by numerical viscosity.

\bibliography{x24}{}

\begin{thebibliography}{}
\expandafter\ifx\csname natexlab\endcsname\relax\def\natexlab#1{#1}\fi
\providecommand{\url}[1]{\href{#1}{#1}}
\providecommand{\dodoi}[1]{doi:~\href{http://doi.org/#1}{\nolinkurl{#1}}}
\providecommand{\doeprint}[1]{\href{http://ascl.net/#1}{\nolinkurl{http://ascl.net/#1}}}
\providecommand{\doarXiv}[1]{\href{https://arxiv.org/abs/#1}{\nolinkurl{https://arxiv.org/abs/#1}}}

\bibitem[{{Adams} \& {Lin}(1993)}]{AdamsLin93}
{Adams}, F.~C., \& {Lin}, D.~N.~C. 1993, in Protostars and Planets III, ed.
  E.~H. {Levy} \& J.~I. {Lunine}, 721

\bibitem[{{Andersen} {et~al.}(2019){Andersen}, {Stephens}, {Dunham}, {Pokhrel},
  {J{\o}rgensen}, {Frimann}, {Segura-Cox}, {Myers}, {Bourke}, {Tobin}, \&
  {Tychoniec}}]{Andersen+19}
{Andersen}, B.~C., {Stephens}, I.~W., {Dunham}, M.~M., {et~al.} 2019, \apj,
  873, 54, \dodoi{10.3847/1538-4357/ab05c7}

\bibitem[{{Baehr} {et~al.}(2017){Baehr}, {Klahr}, \& {Kratter}}]{Baehr+17}
{Baehr}, H., {Klahr}, H., \& {Kratter}, K.~M. 2017, \apj, 848, 40,
  \dodoi{10.3847/1538-4357/aa8a66}

\bibitem[{{Baehr} {et~al.}(2022){Baehr}, {Zhu}, \& {Yang}}]{Baehr+22}
{Baehr}, H., {Zhu}, Z., \& {Yang}, C.-C. 2022, \apj, 933, 100,
  \dodoi{10.3847/1538-4357/ac7228}

\bibitem[{{Balbus} \& {Papaloizou}(1999)}]{BalbusPapaloizou99}
{Balbus}, S.~A., \& {Papaloizou}, J. C.~B. 1999, \apj, 521, 650,
  \dodoi{10.1086/307594}

\bibitem[{{B{\'e}thune} {et~al.}(2021){B{\'e}thune}, {Latter}, \&
  {Kley}}]{Bethune+21}
{B{\'e}thune}, W., {Latter}, H., \& {Kley}, W. 2021, \aap, 650, A49,
  \dodoi{10.1051/0004-6361/202040094}

\bibitem[{{Boley} {et~al.}(2006){Boley}, {Mej{\'\i}a}, {Durisen}, {Cai},
  {Pickett}, \& {D'Alessio}}]{Boley+06}
{Boley}, A.~C., {Mej{\'\i}a}, A.~C., {Durisen}, R.~H., {et~al.} 2006, \apj,
  651, 517, \dodoi{10.1086/507478}

\bibitem[{{Booth} \& {Ilee}(2020)}]{BoothIlee20}
{Booth}, A.~S., \& {Ilee}, J.~D. 2020, \mnras, 493, L108,
  \dodoi{10.1093/mnrasl/slaa014}

\bibitem[{{Booth} {et~al.}(2019){Booth}, {Walsh}, {Ilee}, {Notsu}, {Qi},
  {Nomura}, \& {Akiyama}}]{Booth+19}
{Booth}, A.~S., {Walsh}, C., {Ilee}, J.~D., {et~al.} 2019, \apjl, 882, L31,
  \dodoi{10.3847/2041-8213/ab3645}

\bibitem[{{Boss}(2021)}]{Boss21}
{Boss}, A.~P. 2021, \apj, 923, 93, \dodoi{10.3847/1538-4357/ac2e05}

\bibitem[{{Boss}(2023)}]{Boss23}
---. 2023, \apj, 943, 101, \dodoi{10.3847/1538-4357/acaf63}

\bibitem[{{Boss} \& {Kanodia}(2023)}]{BossKanodia23}
{Boss}, A.~P., \& {Kanodia}, S. 2023, \apj, 956, 4,
  \dodoi{10.3847/1538-4357/acf373}

\bibitem[{{Brucy} \& {Hennebelle}(2021)}]{Brucy+21}
{Brucy}, N., \& {Hennebelle}, P. 2021, \mnras, 503, 4192,
  \dodoi{10.1093/mnras/stab738}

\bibitem[{{Burns} {et~al.}(2023){Burns}, {Uno}, {Sakai}, {Blanchard}, {Rosli},
  {Orosz}, {Yonekura}, {Tanabe}, {Sugiyama}, {Hirota}, {Kim}, {Aberfelds},
  {Volvach}, {Bartkiewicz}, {Caratti o Garatti}, {Sobolev}, {Stecklum},
  {Brogan}, {Phillips}, {Ladeyschikov}, {Johnstone}, {Surcis}, {MacLeod},
  {Linz}, {Chibueze}, {Brand}, {Eisl{\"o}ffel}, {Hyland}, {Uscanga}, {Olech},
  {Durjasz}, {Bayandina}, {Breen}, {Ellingsen}, {van den Heever}, {Hunter}, \&
  {Chen}}]{Burns+23}
{Burns}, R.~A., {Uno}, Y., {Sakai}, N., {et~al.} 2023, Nature Astronomy, 7,
  557, \dodoi{10.1038/s41550-023-01899-w}

\bibitem[{{Cadman} {et~al.}(2021){Cadman}, {Rice}, \& {Hall}}]{Cadman+21}
{Cadman}, J., {Rice}, K., \& {Hall}, C. 2021, \mnras, 504, 2877,
  \dodoi{10.1093/mnras/stab905}

\bibitem[{{Chen} {et~al.}(2023){Chen}, {Jiang}, {Goodman}, \&
  {Ostriker}}]{Chen+23}
{Chen}, Y.-X., {Jiang}, Y.-F., {Goodman}, J., \& {Ostriker}, E.~C. 2023, \apj,
  948, 120, \dodoi{10.3847/1538-4357/acc023}

\bibitem[{{Clarke}(2009)}]{Clarke09}
{Clarke}, C.~J. 2009, \mnras, 396, 1066,
  \dodoi{10.1111/j.1365-2966.2009.14774.x}

\bibitem[{{Cossins} {et~al.}(2009){Cossins}, {Lodato}, \&
  {Clarke}}]{Cossins+09}
{Cossins}, P., {Lodato}, G., \& {Clarke}, C.~J. 2009, \mnras, 393, 1157,
  \dodoi{10.1111/j.1365-2966.2008.14275.x}

\bibitem[{{Currie} {et~al.}(2022){Currie}, {Lawson}, {Schneider}, {Lyra},
  {Wisniewski}, {Grady}, {Guyon}, {Tamura}, {Kotani}, {Kawahara}, {Brandt},
  {Uyama}, {Muto}, {Dong}, {Kudo}, {Hashimoto}, {Fukagawa}, {Wagner}, {Lozi},
  {Chilcote}, {Tobin}, {Groff}, {Ward-Duong}, {Januszewski}, {Norris},
  {Tuthill}, {van der Marel}, {Sitko}, {Deo}, {Vievard}, {Jovanovic},
  {Martinache}, \& {Skaf}}]{Currie+22}
{Currie}, T., {Lawson}, K., {Schneider}, G., {et~al.} 2022, Nature Astronomy,
  6, 751, \dodoi{10.1038/s41550-022-01634-x}

\bibitem[{{Deng} {et~al.}(2021){Deng}, {Mayer}, \& {Helled}}]{Deng+21}
{Deng}, H., {Mayer}, L., \& {Helled}, R. 2021, Nature Astronomy, 5, 440,
  \dodoi{10.1038/s41550-020-01297-6}

\bibitem[{{Deng} {et~al.}(2017){Deng}, {Mayer}, \& {Meru}}]{Deng+17}
{Deng}, H., {Mayer}, L., \& {Meru}, F. 2017, \apj, 847, 43,
  \dodoi{10.3847/1538-4357/aa872b}

\bibitem[{{Drass} {et~al.}(2016){Drass}, {Haas}, {Chini}, {Bayo}, {Hackstein},
  {Hoffmeister}, {Godoy}, \& {Vogt}}]{Drass+16}
{Drass}, H., {Haas}, M., {Chini}, R., {et~al.} 2016, \mnras, 461, 1734,
  \dodoi{10.1093/mnras/stw1094}

\bibitem[{{Galv{\'a}n-Madrid} {et~al.}(2018){Galv{\'a}n-Madrid}, {Liu},
  {Izquierdo}, {Miotello}, {Zhao}, {Carrasco-Gonz{\'a}lez}, {Lizano}, \&
  {Rodr{\'\i}guez}}]{GalvanMadrid+18}
{Galv{\'a}n-Madrid}, R., {Liu}, H.~B., {Izquierdo}, A.~F., {et~al.} 2018, \apj,
  868, 39, \dodoi{10.3847/1538-4357/aae779}

\bibitem[{{Gammie}(2001)}]{Gammie01}
{Gammie}, C.~F. 2001, \apj, 553, 174, \dodoi{10.1086/320631}

\bibitem[{{Gibbons} {et~al.}(2014){Gibbons}, {Mamatsashvili}, \&
  {Rice}}]{Gibbons+14}
{Gibbons}, P.~G., {Mamatsashvili}, G.~R., \& {Rice}, W.~K.~M. 2014, \mnras,
  442, 361, \dodoi{10.1093/mnras/stu809}

\bibitem[{{Gibbons} {et~al.}(2012){Gibbons}, {Rice}, \&
  {Mamatsashvili}}]{Gibbons+12}
{Gibbons}, P.~G., {Rice}, W.~K.~M., \& {Mamatsashvili}, G.~R. 2012, \mnras,
  426, 1444, \dodoi{10.1111/j.1365-2966.2012.21731.x}

\bibitem[{{Hall} {et~al.}(2017){Hall}, {Forgan}, \& {Rice}}]{Hall+17}
{Hall}, C., {Forgan}, D., \& {Rice}, K. 2017, \mnras, 470, 2517,
  \dodoi{10.1093/mnras/stx1244}

\bibitem[{{Hirose} \& {Shi}(2019)}]{HiroseShi19}
{Hirose}, S., \& {Shi}, J.-M. 2019, \mnras, 485, 266,
  \dodoi{10.1093/mnras/stz163}

\bibitem[{{Hopkins} \& {Christiansen}(2013)}]{HopkinsChristiansen13}
{Hopkins}, P.~F., \& {Christiansen}, J.~L. 2013, \apj, 776, 48,
  \dodoi{10.1088/0004-637X/776/1/48}

\bibitem[{{Huang} {et~al.}(2018){Huang}, {Andrews}, {P{\'e}rez}, {Zhu},
  {Dullemond}, {Isella}, {Benisty}, {Bai}, {Birnstiel}, {Carpenter},
  {Guzm{\'a}n}, {Hughes}, {{\"O}berg}, {Ricci}, {Wilner}, \&
  {Zhang}}]{Huang+18}
{Huang}, J., {Andrews}, S.~M., {P{\'e}rez}, L.~M., {et~al.} 2018, \apjl, 869,
  L43, \dodoi{10.3847/2041-8213/aaf7a0}

\bibitem[{{Huang} {et~al.}(2020){Huang}, {Andrews}, {{\"O}berg}, {Ansdell},
  {Benisty}, {Carpenter}, {Isella}, {P{\'e}rez}, {Ricci}, {Williams}, {Wilner},
  \& {Zhu}}]{Huang+20}
{Huang}, J., {Andrews}, S.~M., {{\"O}berg}, K.~I., {et~al.} 2020, \apj, 898,
  140, \dodoi{10.3847/1538-4357/aba1e1}

\bibitem[{{Jiang}(2021)}]{Jiang21}
{Jiang}, Y.-F. 2021, \apjs, 253, 49, \dodoi{10.3847/1538-4365/abe303}

\bibitem[{{Jiang}(2022)}]{Jiang22}
---. 2022, \apjs, 263, 4, \dodoi{10.3847/1538-4365/ac9231}

\bibitem[{{Kratter} \& {Lodato}(2016)}]{KratterLodato16}
{Kratter}, K., \& {Lodato}, G. 2016, \araa, 54, 271,
  \dodoi{10.1146/annurev-astro-081915-023307}

\bibitem[{{Kratter} {et~al.}(2010){Kratter}, {Murray-Clay}, \&
  {Youdin}}]{Kratter+10}
{Kratter}, K.~M., {Murray-Clay}, R.~A., \& {Youdin}, A.~N. 2010, \apj, 710,
  1375, \dodoi{10.1088/0004-637X/710/2/1375}

\bibitem[{{Kubli} {et~al.}(2023){Kubli}, {Mayer}, \& {Deng}}]{Kubli+23}
{Kubli}, N., {Mayer}, L., \& {Deng}, H. 2023, \mnras, 525, 2731,
  \dodoi{10.1093/mnras/stad2478}

\bibitem[{{Laughlin} \& {Rozyczka}(1996)}]{LaughlinRozyczka96}
{Laughlin}, G., \& {Rozyczka}, M. 1996, \apj, 456, 279, \dodoi{10.1086/176648}

\bibitem[{{Lee} {et~al.}(2020){Lee}, {Li}, \& {Turner}}]{Lee+20}
{Lee}, C.-F., {Li}, Z.-Y., \& {Turner}, N.~J. 2020, Nature Astronomy, 4, 142,
  \dodoi{10.1038/s41550-019-0905-x}

\bibitem[{{Lin} \& {Shu}(1966)}]{LinShu66}
{Lin}, C.~C., \& {Shu}, F.~H. 1966, Proceedings of the National Academy of
  Science, 55, 229, \dodoi{10.1073/pnas.55.2.229}

\bibitem[{{Lin} \& {Pringle}(1987)}]{LinPringle87}
{Lin}, D.~N.~C., \& {Pringle}, J.~E. 1987, \mnras, 225, 607,
  \dodoi{10.1093/mnras/225.3.607}

\bibitem[{{Liu}(2020)}]{Liu20}
{Liu}, H.~B. 2020, arXiv e-prints, arXiv:2010.05392.
\newblock \doarXiv{2010.05392}

\bibitem[{{Lodato}(2007)}]{Lodato07}
{Lodato}, G. 2007, Nuovo Cimento Rivista Serie, 30, 293,
  \dodoi{10.1393/ncr/i2007-10022-x}

\bibitem[{{Lodato} \& {Rice}(2004)}]{LodatoRice04}
{Lodato}, G., \& {Rice}, W.~K.~M. 2004, \mnras, 351, 630,
  \dodoi{10.1111/j.1365-2966.2004.07811.x}

\bibitem[{{Lodato} \& {Rice}(2005)}]{LodatoRice05}
---. 2005, \mnras, 358, 1489, \dodoi{10.1111/j.1365-2966.2005.08875.x}

\bibitem[{{Lodato} {et~al.}(2023){Lodato}, {Rampinelli}, {Viscardi},
  {Longarini}, {Izquierdo}, {Paneque-Carre{\~n}o}, {Testi}, {Facchini},
  {Miotello}, {Veronesi}, \& {Hall}}]{Lodato+23}
{Lodato}, G., {Rampinelli}, L., {Viscardi}, E., {et~al.} 2023, \mnras, 518,
  4481, \dodoi{10.1093/mnras/stac3223}

\bibitem[{{Mercer} \& {Stamatellos}(2017)}]{MercerStamatellos17}
{Mercer}, A., \& {Stamatellos}, D. 2017, \mnras, 465, 2,
  \dodoi{10.1093/mnras/stw2714}

\bibitem[{{Meru} {et~al.}(2017){Meru}, {Juh{\'a}sz}, {Ilee}, {Clarke},
  {Rosotti}, \& {Booth}}]{Meru+17}
{Meru}, F., {Juh{\'a}sz}, A., {Ilee}, J.~D., {et~al.} 2017, \apjl, 839, L24,
  \dodoi{10.3847/2041-8213/aa6837}

\bibitem[{{Mullen} {et~al.}(2021){Mullen}, {Hanawa}, \& {Gammie}}]{Mullen+21}
{Mullen}, P.~D., {Hanawa}, T., \& {Gammie}, C.~F. 2021, \apjs, 252, 30,
  \dodoi{10.3847/1538-4365/abcfbd}

\bibitem[{{Nayakshin}(2010)}]{Nayakshin10}
{Nayakshin}, S. 2010, \mnras, 408, L36,
  \dodoi{10.1111/j.1745-3933.2010.00923.x}

\bibitem[{{Nero} \& {Bjorkman}(2009)}]{NeroBjorkman09}
{Nero}, D., \& {Bjorkman}, J.~E. 2009, \apjl, 702, L163,
  \dodoi{10.1088/0004-637X/702/2/L163}

\bibitem[{{Nielsen} {et~al.}(2019){Nielsen}, {De Rosa}, {Macintosh}, {Wang},
  {Ruffio}, {Chiang}, {Marley}, {Saumon}, {Savransky}, {Ammons}, {Bailey},
  {Barman}, {Blain}, {Bulger}, {Burrows}, {Chilcote}, {Cotten}, {Czekala},
  {Doyon}, {Duch{\^e}ne}, {Esposito}, {Fabrycky}, {Fitzgerald}, {Follette},
  {Fortney}, {Gerard}, {Goodsell}, {Graham}, {Greenbaum}, {Hibon}, {Hinkley},
  {Hirsch}, {Hom}, {Hung}, {Dawson}, {Ingraham}, {Kalas}, {Konopacky},
  {Larkin}, {Lee}, {Lin}, {Maire}, {Marchis}, {Marois}, {Metchev},
  {Millar-Blanchaer}, {Morzinski}, {Oppenheimer}, {Palmer}, {Patience},
  {Perrin}, {Poyneer}, {Pueyo}, {Rafikov}, {Rajan}, {Rameau}, {Rantakyr{\"o}},
  {Ren}, {Schneider}, {Sivaramakrishnan}, {Song}, {Soummer}, {Tallis},
  {Thomas}, {Ward-Duong}, \& {Wolff}}]{Nielsen+19}
{Nielsen}, E.~L., {De Rosa}, R.~J., {Macintosh}, B., {et~al.} 2019, \aj, 158,
  13, \dodoi{10.3847/1538-3881/ab16e9}

\bibitem[{{Offner} {et~al.}(2023){Offner}, {Moe}, {Kratter}, {Sadavoy},
  {Jensen}, \& {Tobin}}]{Offner+23}
{Offner}, S.~S.~R., {Moe}, M., {Kratter}, K.~M., {et~al.} 2023, in Astronomical
  Society of the Pacific Conference Series, Vol. 534, Protostars and Planets
  VII, ed. S.~{Inutsuka}, Y.~{Aikawa}, T.~{Muto}, K.~{Tomida}, \& M.~{Tamura},
  275, \dodoi{10.48550/arXiv.2203.10066}

\bibitem[{{Paardekooper} {et~al.}(2023){Paardekooper}, {Dong}, {Duffell},
  {Fung}, {Masset}, {Ogilvie}, \& {Tanaka}}]{Paardekooper+23}
{Paardekooper}, S., {Dong}, R., {Duffell}, P., {et~al.} 2023, in Astronomical
  Society of the Pacific Conference Series, Vol. 534, Protostars and Planets
  VII, ed. S.~{Inutsuka}, Y.~{Aikawa}, T.~{Muto}, K.~{Tomida}, \& M.~{Tamura},
  685, \dodoi{10.48550/arXiv.2203.09595}

\bibitem[{{Paardekooper}(2012)}]{Paardekooper12}
{Paardekooper}, S.-J. 2012, \mnras, 421, 3286,
  \dodoi{10.1111/j.1365-2966.2012.20553.x}

\bibitem[{{Paardekooper} {et~al.}(2011){Paardekooper}, {Baruteau}, \&
  {Meru}}]{Paardekooper+11}
{Paardekooper}, S.-J., {Baruteau}, C., \& {Meru}, F. 2011, \mnras, 416, L65,
  \dodoi{10.1111/j.1745-3933.2011.01099.x}

\bibitem[{{Paneque-Carre{\~n}o} {et~al.}(2021){Paneque-Carre{\~n}o},
  {P{\'e}rez}, {Benisty}, {Hall}, {Veronesi}, {Lodato}, {Sierra}, {Carpenter},
  {Andrews}, {Bae}, {Henning}, {Kwon}, {Linz}, {Loinard}, {Pinte}, {Ricci},
  {Tazzari}, {Testi}, \& {Wilner}}]{Paneque-Carreno+21}
{Paneque-Carre{\~n}o}, T., {P{\'e}rez}, L.~M., {Benisty}, M., {et~al.} 2021,
  \apj, 914, 88, \dodoi{10.3847/1538-4357/abf243}

\bibitem[{{Rafikov}(2005)}]{Rafikov05}
{Rafikov}, R.~R. 2005, \apjl, 621, L69, \dodoi{10.1086/428899}

\bibitem[{{Rice} {et~al.}(2015){Rice}, {Lopez}, {Forgan}, \&
  {Biller}}]{Rice+15}
{Rice}, K., {Lopez}, E., {Forgan}, D., \& {Biller}, B. 2015, \mnras, 454, 1940,
  \dodoi{10.1093/mnras/stv1997}

\bibitem[{{Rice} \& {Armitage}(2009)}]{RiceArmitage09}
{Rice}, W.~K.~M., \& {Armitage}, P.~J. 2009, \mnras, 396, 2228,
  \dodoi{10.1111/j.1365-2966.2009.14879.x}

\bibitem[{{Rice} {et~al.}(2005){Rice}, {Lodato}, \& {Armitage}}]{Rice+05}
{Rice}, W.~K.~M., {Lodato}, G., \& {Armitage}, P.~J. 2005, \mnras, 364, L56,
  \dodoi{10.1111/j.1745-3933.2005.00105.x}

\bibitem[{{Rice} {et~al.}(2004){Rice}, {Lodato}, {Pringle}, {Armitage}, \&
  {Bonnell}}]{Rice+04}
{Rice}, W.~K.~M., {Lodato}, G., {Pringle}, J.~E., {Armitage}, P.~J., \&
  {Bonnell}, I.~A. 2004, \mnras, 355, 543,
  \dodoi{10.1111/j.1365-2966.2004.08339.x}

\bibitem[{{Rice} {et~al.}(2006){Rice}, {Lodato}, {Pringle}, {Armitage}, \&
  {Bonnell}}]{Rice+06}
---. 2006, \mnras, 372, L9, \dodoi{10.1111/j.1745-3933.2006.00215.x}

\bibitem[{{Rice} {et~al.}(2014){Rice}, {Paardekooper}, {Forgan}, \&
  {Armitage}}]{Rice+14}
{Rice}, W.~K.~M., {Paardekooper}, S.~J., {Forgan}, D.~H., \& {Armitage}, P.~J.
  2014, \mnras, 438, 1593, \dodoi{10.1093/mnras/stt2297}

\bibitem[{{Rowther} {et~al.}(2024){Rowther}, {Nealon}, {Meru}, {Wurster},
  {Aly}, {Alexander}, {Rice}, \& {Booth}}]{Rowther+24}
{Rowther}, S., {Nealon}, R., {Meru}, F., {et~al.} 2024, \mnras, 528, 2490,
  \dodoi{10.1093/mnras/stae167}

\bibitem[{{Santos} {et~al.}(2017){Santos}, {Adibekyan}, {Figueira},
  {Andreasen}, {Barros}, {Delgado-Mena}, {Demangeon}, {Faria}, {Oshagh},
  {Sousa}, {Viana}, \& {Ferreira}}]{Santos+17}
{Santos}, N.~C., {Adibekyan}, V., {Figueira}, P., {et~al.} 2017, \aap, 603,
  A30, \dodoi{10.1051/0004-6361/201730761}

\bibitem[{{Schlaufman}(2018)}]{Schlaufman18}
{Schlaufman}, K.~C. 2018, \apj, 853, 37, \dodoi{10.3847/1538-4357/aa961c}

\bibitem[{{Sharma} {et~al.}(2020){Sharma}, {Tobin}, {Sheehan}, {Megeath},
  {Fischer}, {J{\o}rgensen}, {Safron}, \& {Nagy}}]{Sharma+20}
{Sharma}, R., {Tobin}, J.~J., {Sheehan}, P.~D., {et~al.} 2020, \apj, 904, 78,
  \dodoi{10.3847/1538-4357/abbdf4}

\bibitem[{{Shu}(2016)}]{Shu16}
{Shu}, F.~H. 2016, \araa, 54, 667, \dodoi{10.1146/annurev-astro-081915-023426}

\bibitem[{{Stamatellos} {et~al.}(2007){Stamatellos}, {Hubber}, \&
  {Whitworth}}]{Stamatellos+07}
{Stamatellos}, D., {Hubber}, D.~A., \& {Whitworth}, A.~P. 2007, \mnras, 382,
  L30, \dodoi{10.1111/j.1745-3933.2007.00383.x}

\bibitem[{{Stamatellos} \& {Whitworth}(2009)}]{StamatellosWhitworth09}
{Stamatellos}, D., \& {Whitworth}, A.~P. 2009, \mnras, 392, 413,
  \dodoi{10.1111/j.1365-2966.2008.14069.x}

\bibitem[{{Steiman-Cameron} {et~al.}(2023){Steiman-Cameron}, {Durisen},
  {Boley}, {Michael}, {Desai}, \& {McConnell}}]{Steiman-Cameron+23}
{Steiman-Cameron}, T.~Y., {Durisen}, R.~H., {Boley}, A.~C., {et~al.} 2023,
  \apj, 958, 139, \dodoi{10.3847/1538-4357/acff6d}

\bibitem[{Stone {et~al.}(2020)Stone, Tomida, White, \& Felker}]{Stone+20}
Stone, J.~M., Tomida, K., White, C.~J., \& Felker, K.~G. 2020, The
  Astrophysical Journal Supplement Series, 249, 4,
  \dodoi{10.3847/1538-4365/ab929b}

\bibitem[{{Su{\'a}rez} {et~al.}(2019){Su{\'a}rez}, {Downes},
  {Rom{\'a}n-Z{\'u}{\~n}iga}, {Cervi{\~n}o}, {Brice{\~n}o}, {Petr-Gotzens}, \&
  {Vivas}}]{Suarez+19}
{Su{\'a}rez}, G., {Downes}, J.~J., {Rom{\'a}n-Z{\'u}{\~n}iga}, C., {et~al.}
  2019, \mnras, 486, 1718, \dodoi{10.1093/mnras/stz756}

\bibitem[{{Tobin} {et~al.}(2020){Tobin}, {Sheehan}, {Megeath},
  {D{\'\i}az-Rodr{\'\i}guez}, {Offner}, {Murillo}, {van 't Hoff}, {van
  Dishoeck}, {Osorio}, {Anglada}, {Furlan}, {Stutz}, {Reynolds}, {Karnath},
  {Fischer}, {Persson}, {Looney}, {Li}, {Stephens}, {Chandler}, {Cox},
  {Dunham}, {Tychoniec}, {Kama}, {Kratter}, {Kounkel}, {Mazur}, {Maud},
  {Patel}, {Perez}, {Sadavoy}, {Segura-Cox}, {Sharma}, {Stephenson}, {Watson},
  \& {Wyrowski}}]{Tobin+20}
{Tobin}, J.~J., {Sheehan}, P.~D., {Megeath}, S.~T., {et~al.} 2020, \apj, 890,
  130, \dodoi{10.3847/1538-4357/ab6f64}

\bibitem[{{Tsukamoto} {et~al.}(2023){Tsukamoto}, {Maury}, {Commercon}, {Alves},
  {Cox}, {Sakai}, {Ray}, {Zhao}, \& {Machida}}]{Tsukamoto+23}
{Tsukamoto}, Y., {Maury}, A., {Commercon}, B., {et~al.} 2023, in Astronomical
  Society of the Pacific Conference Series, Vol. 534, Protostars and Planets
  VII, ed. S.~{Inutsuka}, Y.~{Aikawa}, T.~{Muto}, K.~{Tomida}, \& M.~{Tamura},
  317

\bibitem[{{Tychoniec} {et~al.}(2018){Tychoniec}, {Tobin}, {Karska}, {Chandler},
  {Dunham}, {Harris}, {Kratter}, {Li}, {Looney}, {Melis}, {P{\'e}rez},
  {Sadavoy}, {Segura-Cox}, \& {van Dishoeck}}]{Tychoniec+18}
{Tychoniec}, {\L}., {Tobin}, J.~J., {Karska}, A., {et~al.} 2018, \apjs, 238,
  19, \dodoi{10.3847/1538-4365/aaceae}

\bibitem[{{Veronesi} {et~al.}(2021){Veronesi}, {Paneque-Carre{\~n}o}, {Lodato},
  {Testi}, {P{\'e}rez}, {Bertin}, \& {Hall}}]{Veronesi+21}
{Veronesi}, B., {Paneque-Carre{\~n}o}, T., {Lodato}, G., {et~al.} 2021, \apjl,
  914, L27, \dodoi{10.3847/2041-8213/abfe6a}

\bibitem[{{Vigan} {et~al.}(2021){Vigan}, {Fontanive}, {Meyer}, {Biller},
  {Bonavita}, {Feldt}, {Desidera}, {Marleau}, {Emsenhuber}, {Galicher}, {Rice},
  {Forgan}, {Mordasini}, {Gratton}, {Le Coroller}, {Maire}, {Cantalloube},
  {Chauvin}, {Cheetham}, {Hagelberg}, {Lagrange}, {Langlois}, {Bonnefoy},
  {Beuzit}, {Boccaletti}, {D'Orazi}, {Delorme}, {Dominik}, {Henning}, {Janson},
  {Lagadec}, {Lazzoni}, {Ligi}, {Menard}, {Mesa}, {Messina}, {Moutou},
  {M{\"u}ller}, {Perrot}, {Samland}, {Schmid}, {Schmidt}, {Sissa}, {Turatto},
  {Udry}, {Zurlo}, {Abe}, {Antichi}, {Asensio-Torres}, {Baruffolo}, {Baudoz},
  {Baudrand}, {Bazzon}, {Blanchard}, {Bohn}, {Brown Sevilla}, {Carbillet},
  {Carle}, {Cascone}, {Charton}, {Claudi}, {Costille}, {De Caprio},
  {Delboulb{\'e}}, {Dohlen}, {Engler}, {Fantinel}, {Feautrier}, {Fusco},
  {Gigan}, {Girard}, {Giro}, {Gisler}, {Gluck}, {Gry}, {Hubin}, {Hugot},
  {Jaquet}, {Kasper}, {Le Mignant}, {Llored}, {Madec}, {Magnard}, {Martinez},
  {Maurel}, {M{\"o}ller-Nilsson}, {Mouillet}, {Moulin}, {Orign{\'e}}, {Pavlov},
  {Perret}, {Petit}, {Pragt}, {Puget}, {Rabou}, {Ramos}, {Rickman}, {Rigal},
  {Rochat}, {Roelfsema}, {Rousset}, {Roux}, {Salasnich}, {Sauvage}, {Sevin},
  {Soenke}, {Stadler}, {Suarez}, {Wahhaj}, {Weber}, \& {Wildi}}]{Vigan+21}
{Vigan}, A., {Fontanive}, C., {Meyer}, M., {et~al.} 2021, \aap, 651, A72,
  \dodoi{10.1051/0004-6361/202038107}

\bibitem[{{Vorobyov}(2013)}]{Vorobyov13}
{Vorobyov}, E.~I. 2013, \aap, 552, A129, \dodoi{10.1051/0004-6361/201220601}

\bibitem[{{Vorobyov} {et~al.}(2021){Vorobyov}, {Elbakyan}, {Liu}, \&
  {Takami}}]{Vorobyov+21}
{Vorobyov}, E.~I., {Elbakyan}, V.~G., {Liu}, H.~B., \& {Takami}, M. 2021, \aap,
  647, A44, \dodoi{10.1051/0004-6361/202039391}

\bibitem[{{Whitworth} \& {Stamatellos}(2006)}]{WhitworthStamatellos06}
{Whitworth}, A.~P., \& {Stamatellos}, D. 2006, \aap, 458, 817,
  \dodoi{10.1051/0004-6361:20065806}

\bibitem[{{Wu} {et~al.}(2024){Wu}, {Chen}, \& {Lin}}]{Wu+24}
{Wu}, Y., {Chen}, Y.-X., \& {Lin}, D. N.~C. 2024, \mnras, 528, L127,
  \dodoi{10.1093/mnrasl/slad183}

\bibitem[{{Xu}(2022{\natexlab{a}})}]{Xu22}
{Xu}, W. 2022{\natexlab{a}}, \apj, 934, 156, \dodoi{10.3847/1538-4357/ac7b94}

\bibitem[{{Xu}(2022{\natexlab{b}})}]{XuThesis}
---. 2022{\natexlab{b}}, PhD thesis, Princeton University.
\newblock \url{http://arks.princeton.edu/ark:/88435/dsp018623j196s}

\bibitem[{{Xu} \& {Armitage}(2023)}]{XA23}
{Xu}, W., \& {Armitage}, P.~J. 2023, \apj, 946, 94,
  \dodoi{10.3847/1538-4357/acb7e5}

\bibitem[{{Xu} {et~al.}(2024){Xu}, {Jiang}, {Kunz}, \& {Stone}}]{paper2}
{Xu}, W., {Jiang}, Y.-F., {Kunz}, M.~W., \& {Stone}, J.~M. 2024

\bibitem[{{Xu} \& {Kunz}(2021{\natexlab{a}})}]{XK21a}
{Xu}, W., \& {Kunz}, M.~W. 2021{\natexlab{a}}, \mnras, 502, 4911,
  \dodoi{10.1093/mnras/stab314}

\bibitem[{{Xu} \& {Kunz}(2021{\natexlab{b}})}]{XK21b}
---. 2021{\natexlab{b}}, \mnras, 508, 2142, \dodoi{10.1093/mnras/stab2715}

\bibitem[{{Xu} {et~al.}(2023){Xu}, {Ohashi}, {Aso}, \& {Liu}}]{Xu+23}
{Xu}, W., {Ohashi}, S., {Aso}, Y., \& {Liu}, H.~B. 2023, \apj, 954, 190,
  \dodoi{10.3847/1538-4357/aced4c}

\bibitem[{{Young} \& {Clarke}(2015)}]{YoungClarke15}
{Young}, M.~D., \& {Clarke}, C.~J. 2015, \mnras, 451, 3987,
  \dodoi{10.1093/mnras/stv1266}

\bibitem[{{Young} \& {Clarke}(2016)}]{YoungClarke16}
---. 2016, \mnras, 455, 1438, \dodoi{10.1093/mnras/stv2378}

\bibitem[{{Zhu} {et~al.}(2012){Zhu}, {Hartmann}, {Nelson}, \&
  {Gammie}}]{Zhu+12}
{Zhu}, Z., {Hartmann}, L., {Nelson}, R.~P., \& {Gammie}, C.~F. 2012, \apj, 746,
  110, \dodoi{10.1088/0004-637X/746/1/110}

\end{thebibliography}
\bibliographystyle{aasjournal}

\end{CJK*}
\end{document}